\documentclass[twocolumn]{aastex63}
\usepackage{mathtools}
\usepackage[maxfloats=40]{morefloats} 
\usepackage{color}
\usepackage{amsmath}
\usepackage[figure,figure*]{hypcap}
\DeclarePairedDelimiter\abs{\lvert}{\rvert}%
\DeclarePairedDelimiter\norm{\lVert}{\rVert}%

\makeatletter
\let\oldabs\abs
\def\abs{\@ifstar{\oldabs}{\oldabs*}}
\let\oldnorm\norm
\def\norm{\@ifstar{\oldnorm}{\oldnorm*}}
\makeatother

\newcommand{\Rom}[1]{\uppercase\expandafter{\romannumeral #1\relax}}
\def \kms {km~s$^{-1}$}
\def\C18O{\textrm{C$^{18}$O}}
\def\13CO{\textrm{$^{13}$CO}}
\def\nh3{\textrm{NH$_{3}$}}

\begin{document}

\title{Multi-Scale Magnetic Field Observations Reveal how Colliding Flows Trigger Star Formation}

\author[0000-0002-6668-974X]{Jia-Wei Wang}
\affiliation{East Asian Observatory, 660 N. A'oh\={o}k\={u} Place, University Park, Hilo, HI 96720, USA}

\author[0000-0003-2777-5861]{Patrick M. Koch}
\affiliation{Academia Sinica Institute of Astronomy and Astrophysics, No.1, Sec. 4., Roosevelt Road, Taipei 10617, Taiwan}

\author[0000-0003-2300-2626]{Hauyu Baobab Liu}
\affiliation{Department of Physics, National Sun Yat-Sen University, No. 70, Lien-Hai Road, Kaohsiung City 80424, Taiwan}

\author[0000-0002-5714-799X]{Valentin J. M. Le Gouellec}
\affiliation{Institut de Cienci\`es de l’Espai (ICE-CSIC), Campus UAB, Carrer de Can Magrans S/N, E-08193 Cerdanyola del Vall\`es, Spain}
\affiliation{Institut d’Estudis Espacials de Catalunya (IEEC), c/ Gran Capitá, 2-4, 08034 Barcelona, Spain}

\author[0000-0001-9299-5479]{Yuxin Lin}
\affiliation{Max-Planck-Institut f\"ur Extraterrestrische Physik, Giessenbachstr. 1, D-85748 Garching bei M\"unchen, Germany}

\author[0000-0003-2384-6589]{Qizhou Zhang}
\affiliation{Center for Astrophysics | Harvard \& Smithsonian, 60 Garden Street, Cambridge, MA 02138, USA}

\author[0000-0001-5522-486X]{Shih-Ping Lai}
\affiliation{Institute of Astronomy and Department of Physics, National Tsing Hua University, Hsinchu 30013, Taiwan}

\begin{abstract}
Magnetic fields play a crucial yet complex role in star formation, while their connection between large-scale filamentary clouds and small-scale young stellar objects remains poorly understood. We present new continuum polarization observations from the JCMT, ACA, and ALMA that provide continuous magnetic field measurements from $\sim$5 pc down to $\sim$4000 au, tracing for the first time the evolution of field morphology seamlessly across all key scales within a massive star-forming system. Our polarization maps reveal multiple U-shaped magnetic field structures pointing toward the central protocluster, aligned with accreting filaments from parsec to subparsec-scales and converging at the compact center. This morphology suggests an environment of colliding flows that drag magnetic fields and trigger massive protocluster formation. On $\sim$4000 au scales, we identify compact U-shaped fields likely guiding the kinematics of streamers accreting onto dense cores. The increasing curvature of these U-shaped patterns is a direct measure of a growing magnetic field tension force, implying a magnetic field strength scaling index of $0.50\pm0.10$. These results indicate that the field, possibly enhanced by large-scale flow collisions, becomes strong enough to regulate star formation, linking parsec-scale colliding flows, a subparsec hub-filament system, and the triggering of massive star formation.

\end{abstract}


\section{Introduction}
Star formation is a multiscale process covering an astonishing range in physical length of 6 or 7 orders of magnitude: it is initiated by the kinematics of giant molecular clouds on scales beyond 10 parsecs, develops within parsec-scale interstellar filaments, fragments into dense cores on subparsec-scales, condenses into young stellar objects on $\sim$1000-au scales, and ultimately forms stars on au scales. Magnetic fields (hereafter B-fields) are predicted to play a crucial role throughout this hierarchy by opposing gravity and guiding the direction of collapse \citep{cr12,pl16,an14,li14}. Yet their exact role across scales remains poorly understood, largely because of observational limitations.

The JCMT B-fields In STar-forming Region Observations (BISTRO) survey \citep{wa17} has shown that B-fields are often well organized and frequently aligned or anti-aligned with interstellar filaments on (sub)parsec scales \citep[e.g.,][]{do20,ar20,kw22}, pointing to a close coupling between magnetic and filamentary structures \citep{pa23}. However, estimates of the mass-to-flux ratio span a wide range, from subcritical to supercritical \citep[e.g.,][]{pa17,hw21}, implying that the dynamical importance of B-fields is strongly environment-dependent. On smaller au scales, high-resolution ALMA polarization surveys toward young stellar objects have revealed strikingly diverse morphologies, including classical hourglass patterns associated with magnetically regulated collapse, spirals shaped by rotating infall, and disordered structures consistent with turbulence-dominated conditions \citep[e.g.,][]{hu19,ma22,fe21,hu25}.

Together, these findings highlight the existence of a missing intermediate regime linking the relatively ordered magnetic fields observed on parsec scales with the complex and chaotic morphologies seen on $\lesssim 0.1$ pc scales in dense cores and envelopes. Probing this regime has long been difficult due to the gap between the spatial scales accessible to single-dish telescopes and those resolved by interferometers. As a result, most previous multi-scale studies have been limited to reporting inconsistencies between large- and small-scale fields, leaving the physical mechanisms driving these variations poorly understood. Bridging this gap by mapping magnetic fields continuously across scales is therefore essential for a comprehensive picture of star formation: ordered fields may reveal how kinematics and density structure are regulated by magnetic tension, while complex structures may preserve signatures of energetic events, such as cloud–cloud collisions \citep[e.g.,][]{in13,fr15,wa22}, that reshape magnetic fields during the earliest evolutionary stages.

G33.92+0.11 (hereafter G33) is a remarkable ultracompact H\Rom{2} region located in the Galactic plane at a distance of $7.1\substack{+1.2 \\ -0.3}$ kpc \citep{fi03}. Early studies found that its derived virial masses are significantly smaller than the enclosed molecular gas masses, suggesting that the system is geometrically thin and viewed nearly face-on \citep{wa99,liu12}. Subsequent 1.3 mm continuum observations with the Submillimeter Array (SMA) revealed that G33 consists of parsec-scale filaments converging toward a massive central hub within the inner $\sim$0.6 pc, which contains a gas mass of $3.0\substack{+2.8 \\ -1.4}\times10^3$ M$_{\sun}$ \citep{liu12}. This morphology identifies G33 as a typical hub–filament system.

Higher-resolution observations with ALMA further resolve the inner 0.6 pc region, revealing several spiral-like arms feeding two central massive molecular cores with masses of $100$–$300$ M$_{\sun}$ \citep{liu15}. Molecular line observations show clear velocity gradients of $\sim1~{\rm km~s^{-1}~pc^{-1}}$ on parsec scales, increasing to $\sim5~{\rm km~s^{-1}~pc^{-1}}$ within the innermost 0.6 pc along the spiral-like arms. This trend suggests that the arms trace gravitationally dominated accretion flows, accelerated from ambient infalling material \citep{liu12,liu15}. Based on the ALMA continuum image, \citet{liu19} identified 28 Class 0/I young stellar object candidates distributed both along the spiral-like structures and within the central cores. Most of these sources are associated with SiO jets, indicating active and ongoing massive star formation within the system.

Magnetic-field observations have added another important piece to this picture. Polarization mapping with the James Clerk Maxwell Telescope (JCMT) POL-2 instrument reveals a converging magnetic-field morphology on $\sim$1–10 pc scales toward the central protocluster, aligned with the parsec-scale filamentary structures, velocity gradient, and the inferred gravitational forces \citep{wa21}. These results suggest that the magnetic-field structure reflects the influence of gravitationally driven inflows along the parsec-scale filaments, supporting a scenario in which the central massive protocluster forms within a hub fed by large-scale accretion flows. 

Despite this progress, the connection between the large-scale filaments and the smaller spiral-like structures remains uncertain due to the lack of observations probing the intermediate spatial scales between JCMT and ALMA measurements. In particular, the orientations of the filaments identified on parsec scales and those observed on $\sim$1000-au scales are not fully consistent, raising questions about whether they are dynamically connected. Furthermore, the origin of the angular momentum responsible for the spiral-like structures in the inner hub remains unclear. 

To address this gap, we present new ALMA Band 6 continuum polarization observations of the massive hub–filament system G33, obtained with the Atacama Compact Array (ACA). These observations probe the previously missing intermediate spatial regime between parsec and 1000-au scales. By combining the ACA data with high-resolution ALMA 12-m array observations and archival JCMT 850 $\mu$m polarization maps, we obtain continuous measurements of the magnetic-field morphology from $\sim$10 pc down to $\sim$4000 au. This multi-scale dataset enables us, for the first time, to trace the evolution of the magnetic-field structure across all key spatial scales within a single massive star-forming system, rather than relying on isolated snapshots at different resolutions.

In \autoref{sec:obs}, we describe the observations and data reduction. The observed multi-scale intensity and magnetic-field morphology are presented in \autoref{sec:results}. In \autoref{sec:ana}, we analyze the gas kinematics and magnetic-field properties and discuss the role of magnetic fields across multiple spatial scales. Our conclusions are summarized in \autoref{sec:con}.

\section{Observations}\label{sec:obs}
\subsection{ALMA Observations}\label{sec:pol}
ALMA ACA Band~6 (211 - 275 GHz)
continuum observations toward G33.92+0.11 were carried out from 2023 August 19--21 (project 2022.1.01482.S). The correlator was configured in Frequency Division Mode (FDM) for full-polarization cross correlations. The spectral setup included three spectral windows, each with a resolution of 3.125~MHz (4.33~km~s$^{-1}$) and a bandwidth of 1.875~GHz, to map the dust continuum. An additional spectral window with a resolution of 98~kHz (0.13~km~s$^{-1}$) and a bandwidth of 58.59~MHz was configured for the C$^{18}$O~(2--1) line to trace the gas kinematics. 

ALMA 12-m array Band~6 continuum observations of G33.92+0.11 were conducted from 2024 September 5--14 in C43-3 configuration (project 2023.1.01004.S). The correlator was also set to FDM for full-polarization cross correlations. Four spectral windows were configured with a resolution of 2.258~MHz (3.0~km~s$^{-1}$) and a bandwidth of 1.875~GHz to optimize continuum sensitivity.

The standard calibration was performed by the ALMA Regional Center using CASA version~6.4 for ACA and 6.6.1 for 12-m data. The measured instrumental polarization leakage (D-terms) were in general $<5\%$. To improve image quality, we applied the IMAGER pipeline\footnote{\url{https://imager.oasu.u-bordeaux.fr}} (version~4.5) to the calibrated data. We first extracted Stokes $I$ visibilities from the CASA measurement set. The pipeline was run on the Stokes~$I$ data to identify line contamination and to derive a self-calibration solution from the continuum channels. The resulting solution was applied to both the line channels. Continuum and spectral-line images were then restored from the self-calibrated datasets. The polarization fraction and orientation were computed on a pixel-by-pixel basis. The calculated polarization fraction $P$ was debiased using the asymptotic estimator \citep{wa74} as
\begin{equation}\label{eq:debias}
P=\frac{1}{I}\sqrt{(Q^2+U^2)-\frac{1}{2}(\sigma_{Q}^2+\sigma_{U}^2)}
\end{equation}
with a polarization uncertainty $\sigma_{P}$ calculated as
\begin{equation}\label{eq:dp}
\sigma_{P}=\sqrt{\frac{Q^2\sigma_{Q}^2+U^2\sigma_{U}^2}{(Q^2+U^2)I^2}+\frac{\sigma_{I}^2(Q^2+U^2)}{I^{4}}},
\end{equation}
where $\sigma_{I}$, $\sigma_{Q}$, and $\sigma_{U}$ are the uncertainties in the Stokes 
$I$, $Q$, and $U$. Only polarization detections above $3\sigma_P$ (for ACA) and $2\sigma_P$ (for 12-m) were retained for further analyses.

The polarization position angle ($PA$) is computed as
\begin{equation}\label{eq:PA}
PA=\frac{1}{2}\tan^{-1}(\frac{U}{Q}),
\end{equation}
and the corresponding uncertainty is estimated as
\begin{equation}\label{eq:ePA}
\delta PA=\frac{1}{2}\sqrt{\frac{(Q^2\delta U^2+U^2\delta Q^2)}{(Q^2+U^2)^2}} .
\end{equation}
The magnetic field orientations in this paper are inferred to be $PA +90\degr$.

The final ACA continuum images achieved an angular resolution of $7\farcs4 \times 4\farcs3$ with a position angle of $84^\circ$. The rms noise levels were 1.8~mJy\,beam$^{-1}$ (dynamic-range limited) for Stokes~$I$ and 0.24~mJy\,beam$^{-1}$ for Stokes~$Q$ and $U$. The final ALMA 12-m continuum images achieved an angular resolution of $0\farcs72 \times 0\farcs56$ with a position angle of $-85^\circ$, and rms noise levels of 0.47~mJy\,beam$^{-1}$ (dynamic-range limited) for Stokes~$I$ and 0.02~mJy\,beam$^{-1}$ for Stokes~$Q$ and $U$. Additional data from the ACA C$^{18}$O~(2--1) line, with velocity resolutions of 0.13 \kms was used to trace gas kinematics. The datasets used in this paper are summarized in \autoref{obstab}.

\begin{deluxetable*}{ccccCcc}
\tablecaption{Datasets.\label{obstab}}
\renewcommand{\thetable}{\arabic{table}}
\tablehead{\colhead{Instrument} & \colhead{Line} & \colhead{$\theta_{beam}$} & \colhead{PA$_{beam}$} & \colhead{Velocity Resolution} & \colhead{Maxium Recoverable Scale} & \colhead{$\sigma_{RMS}$} \\
\colhead{} & \colhead{} & \colhead{(arcsec)} & \colhead{(deg)} &\colhead{(km/s)} &\colhead{arcsec} &\colhead{} }
\startdata
\hline
JCMT & Cont. Polarization & 14 & ... & ... & ... & 1.1~mJy/beam \\
ALMA-ACA & Cont. Polarization & $7.4 \times 4.3$ & 84 & ... & 29 &1.8~mJy/beam \\
ALMA-12m & Cont. Polarization & 0.72$\times$0.56 & -85 & ... & 12.4 & 0.47~mJy/beam \\
IRAM-30m & C$^{18}$O (2--1) & 11.8 & ... & 0.066 & ... & 0.2~K \\
JVLA & NH$_{3}$ (1,1) & 4.2$\times$3.7 & -10 & 0.05 & 61 & 2.3~mJy/beam \\
ALMA-ACA & C$^{18}$O (2--1) & $7.2 \times 4.8$ & 87 & 0.08 & 29 & 45~mJy/beam \\
ALMA-12m+ACA & $^{13}$CS (5--4) & 0.18$\times$0.13 & -89 & 0.18 & 29 & 2.3~mJy/beam \\
\enddata
\end{deluxetable*}

\subsection{JVLA NH$_{3}$ Observations}\label{sec:harp}
NH$_{3}$ (1,1) line observations taken with JVLA in the dual polarization (RR, LL) mode with configuration D (Project code: 19B-069) is adopted to recover the extended gas structures between ACA and JCMT. The data were processed using CASA version~5.6.2 with self-calibration using a solution interval of 120 seconds. The final NH$_{3}$ image, with Briggs Robust$=$2 weighted, has a synthesized beam of 4\farcs2$\times$3\farcs7 (P.A.=-10$^{\circ}$) and velocity channel widths of 0.05 km\,s$^{-1}$. The rms noise in each velocity channel is $\sim$2.3 mJy\,beam$^{-1}$ ($\sim$0.29 K at 23 GHz). Only the main hyperfine component (rest frequency 23.6945~GHz) is used in our analysis. 

\subsection{Archival Data}\label{sec:pol2}
To probe parsec-scale structures, we incorporated JCMT POL-2 polarization data (resolution $14^{\prime\prime}$) and IRAM~30-m C$^{18}$O~(2--1) data (resolution $12^{\prime\prime}$) from \citet{wa20}. In addition, $^{13}$CS~(5--4) data combined from ALMA ACA and 12-m observations from \citet{liu15}, with a beam size of $0\farcs18 \times 0\farcs13$ and a spectral resolution of 0.18 \kms, were used to trace gas kinematics at the envelope-scale and to compare with our ALMA 12-m continuum polarization data.

\section{Results}\label{sec:results}
\begin{figure*}
\includegraphics[width=\textwidth]{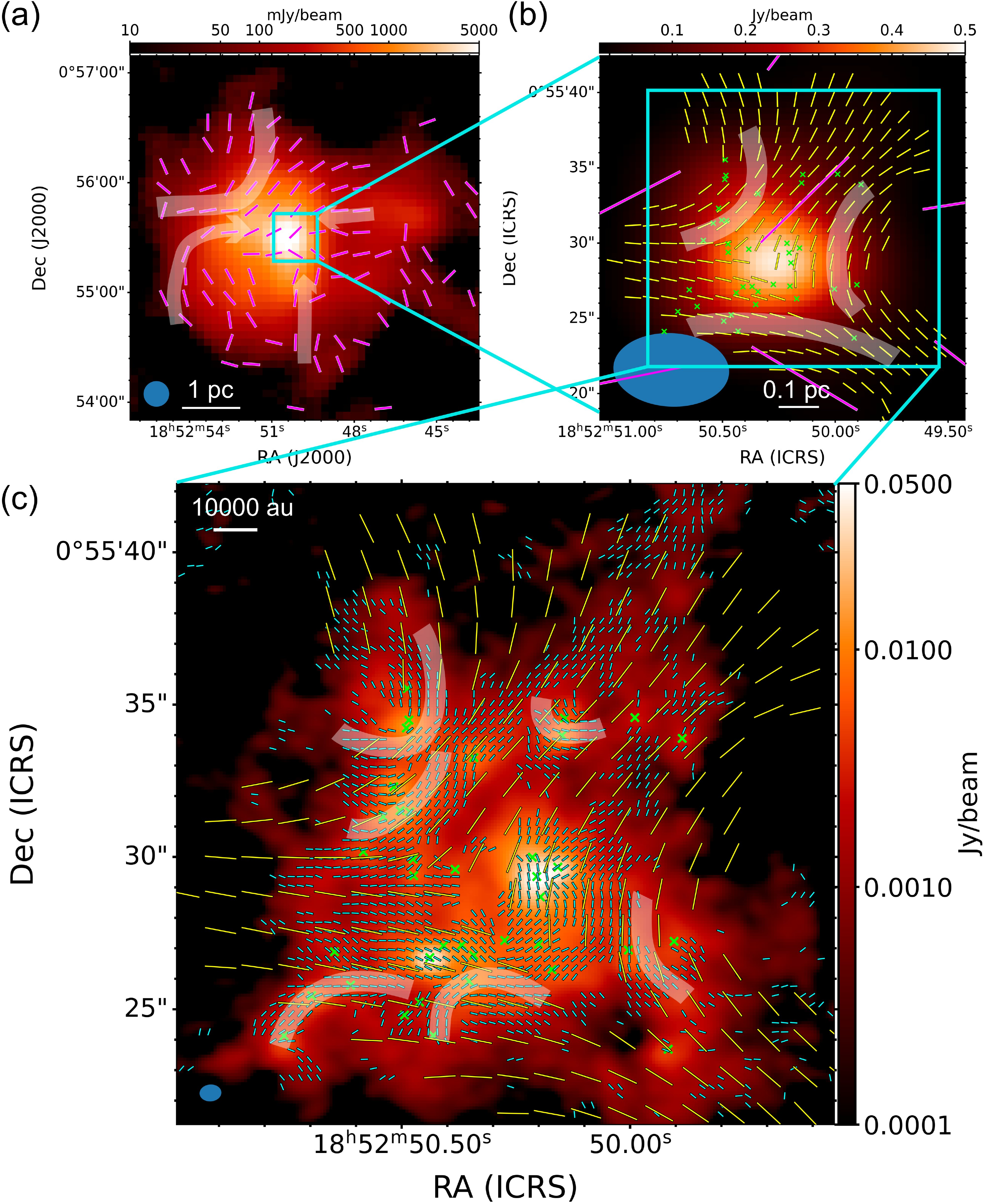}
\caption{Multi-scale magnetic field maps of G33.92+0.11. (a) JCMT polarization segments (magenta) overlaid on the JCMT 850\,$\mu$m continuum image, showing a converging magnetic fields (white arrows) and a U-shape morphology in the Northeast (white curve). (b) ALMA ACA polarization segments (yellow) overlaid on the ACA 1.3\,mm continuum image, showing three U-shape magnetic fields (white curves). (c) ALMA 12-m array polarization segments (cyan) overlaid on the 12-m array 1.3\,mm continuum image. All magnetic field segments are rotated by 90$^\circ$ from the originally detected polarization orientations. Additionally shown are the dense cores identified in \citet{su21} with green crosses. The different telescope resolutions are displayed with the blue ellipses in the lower left corners.}
\label{fig:pmap}
\end{figure*}

G33 is a massive star-forming system that hosts a central protocluster embedded within a dense hub and associated with an ultracompact H\textsc{ii} region \citep{fi03}. JCMT POL-2 observations reveal converging magnetic field lines toward the central protocluster
(white arrows in \autoref{fig:pmap}a), with an U-shape pattern (white curve) located at the northeastern side of the source. \citet{wa21} show that the converging magnetic field orientations are consistent with converging filaments and projected gravitational forces, suggesting that the field morphology reflects the influence of gravitationally driven infalling filaments.

Our new ALMA ACA polarization data, with a synthesized beam of $7.7^{\prime\prime} \times 4.9^{\prime\prime}$, reveal the magnetic field morphology at subparsec-scales, connecting the large-scale converging filaments to the central cluster-forming region (\autoref{fig:pmap}b). The magnetic field orientation map clearly shows three U-shaped structures converging from the northeast, northwest, and south toward the central protocluster. U-shaped magnetic field morphologies are commonly observed in regions influenced by accreting flows or cloud–cloud collisions, where field lines are pinched by flow ram pressure or shocks \citep[e.g.,][]{go18,wa20,wa22}. To reveal how magnetic fields correlate with condensed structures, we adopt the dense cores identified in \citet{su21} from the ALMA ACA and 12-m combined 1.3 mm continuum image with the \textit{astrodendro} package.

The high-resolution ALMA observations, with a synthesized beam of $0.7^{\prime\prime} \times 0.5^{\prime\prime}$, resolve a compact star-forming protocluster surrounded by at least three spiral arms in the innermost region (\autoref{fig:pmap}c). A logarithmic color scale is adopted here to highlight the extended structures, while a linear-scale map emphasizing the spiral-arm ridges is presented in \autoref{sec:3.2}. The spiral morphology suggests that rotational motions play an important role in the local gas dynamics \citep{liu15}, in contrast to the larger-scale regions where such signatures are not clearly evident. This morphological transition implies a change in the dominant dynamical processes between large and small spatial scales. In addition, the polarization map resolves numerous U-shaped magnetic-field features on $\sim$4000 au scales (hereafter envelope scales), which are embedded within the intermediate-scale U-shaped magnetic structures (\autoref{fig:pmap}c). These envelope-scale U-shaped features are generally more tightly curved than their larger-scale counterparts (see further analysis in \autoref{sec:3.1}), though their orientations are not always aligned. 

\section{Analysis and Discussion}\label{sec:ana}
\subsection{Magnetic Field Curvature}\label{sec:3.1}
In order to quantitatively confirm the apparent U-shaped magnetic-field structures shown in \autoref{fig:pmap}, we derived the curvature of the magnetic field from the polarization orientation map.
To quantify the local magnetic-field curvature while properly handling the 0\degr/180\degr ambiguity in polarization orientations, we define a complex orientation field
\begin{equation}
w(x,y) = e^{i\,2\theta(x,y)},
\end{equation}
where $\theta$ is the magnetic-field orientation measured east of north.
The factor two removes the $\theta\equiv\theta+\pi$ ambiguity.

Differentiating $w = e^{i\psi}$ with $\psi=2\theta$ gives
\begin{equation}
\nabla w = i\,w\,\nabla\psi \;\;\Rightarrow\;\; \nabla\psi = \Im\!\left(w^* \nabla w \right),
\end{equation}
where $\Im(\,)$ denotes the imaginary part and $w^*$ is the complex conjugate of $w$.
Hence,
\begin{equation} \nabla\theta = \tfrac{1}{2}\,\Im\!\left(w^* \nabla w\right). \end{equation}

The curvature $\kappa$ of the magnetic-field streamline is defined as the rate of change of orientation angle $\theta$ along its tangent ($\hat{t}$):
\begin{equation} 
\kappa = \left| \frac{d\theta}{ds} \right| = \left| \hat{t}\cdot\nabla\theta \right|, 
\end{equation}
where $\hat{t}=(u,v)=(\sin\theta,\cos\theta)$ and $s$ denotes the arc length.
Combining the above expressions yields
\begin{equation} 
\kappa = \frac{1}{2}\, \Bigl|\, \hat{t}\cdot \Im\!\bigl(w^* \nabla w\bigr) \Bigr|. 
\end{equation}

We computed $\nabla w$ using a circular moving window, fitting a plane $u=ax+by+c$ to the real and imaginary parts of $w$.
Window radii of $72^{\prime\prime}$, $5^{\prime\prime}$, and $0\farcs6$ were used for the JCMT, ACA, and ALMA 12-m data, respectively, corresponding to the scale of the observed U-shaped features.
Curvature values, expressed in arcsec$^{-1}$ after dividing by the pixel scale, are equivalent to
\begin{equation}
\kappa = \bigl| u\,\theta_E + v\,\theta_N \bigr|,
\end{equation}
where $\theta_E=\partial\theta/\partial E$ and $\theta_N=\partial\theta/\partial N$ are the orientation gradients along the east and north directions.
The transformation from image coordinates $(\partial\theta/\partial x$ and $\partial\theta/\partial y)$ to sky coordinates $(E,N)$ was performed using the World Coordinate System Jacobian matrix. Only windows with at least 25 \% of valid polarization detections were used.

The uncertainty in curvature was estimated as
\begin{equation}
    \sigma_\kappa = \sqrt{2}\,\frac{\sigma{_\theta}}{H},
\end{equation}
where $H$ is the diameter of the moving window and $\sigma_{\theta}$ is the median polarization-angle uncertainty of $8.5^\circ$, $3.4^\circ$ and $2.7^\circ$ for the JCMT, ACA, and 12-m data, respectively.  
This yields curvature uncertainties ($\sigma_{\rm curve}$) of 0.0015, 0.008 and 0.055~arcsec$^{-1}$ for the JCMT, ACA, and 12-m maps. The moving-window size and the measurement uncertainty set the upper and lower limits on the curvature detectable in each dataset.

\begin{figure*}
\includegraphics[width=\textwidth]{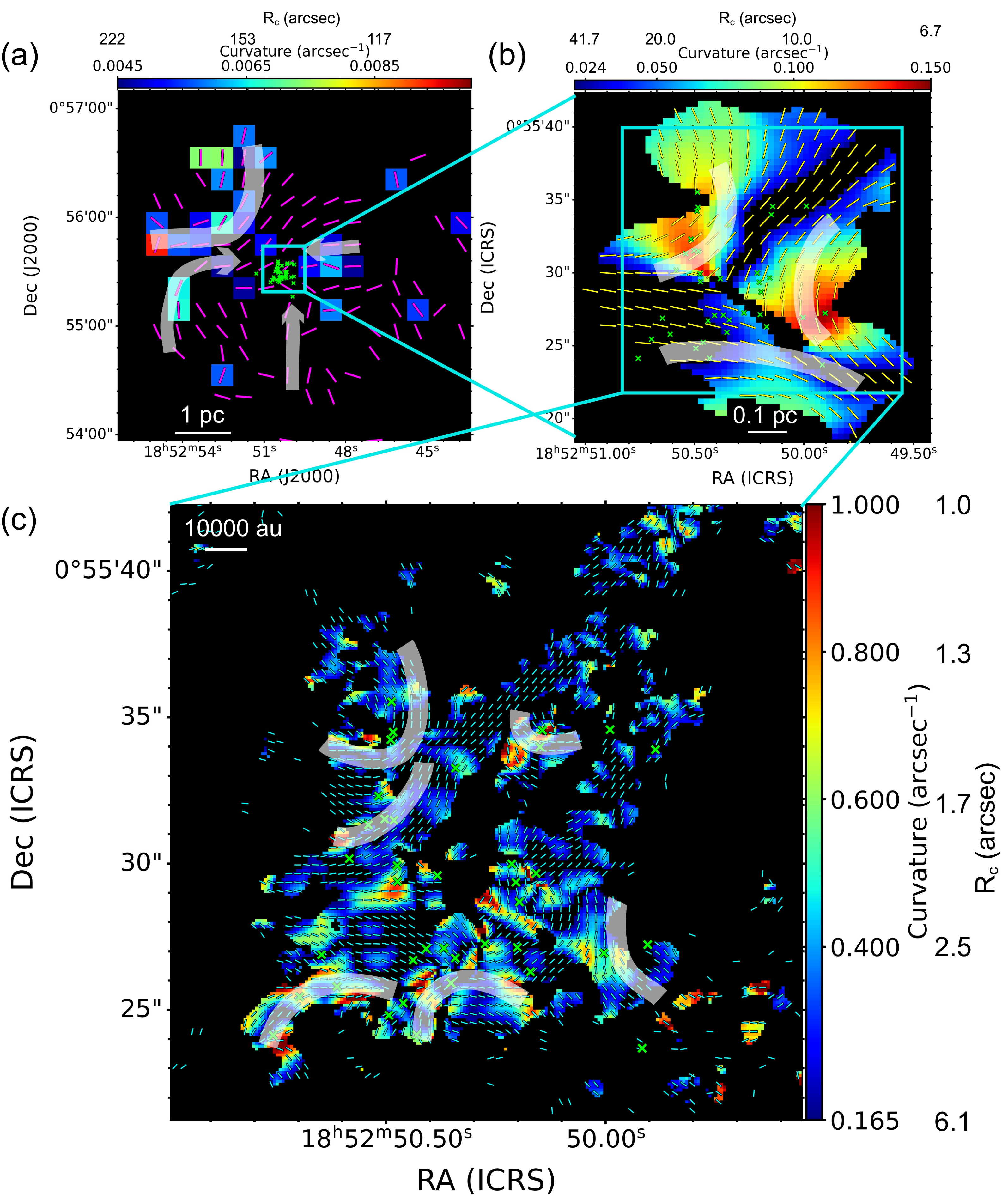}
\caption{Magnetic field curvature ($\kappa$) estimated from (a) JCMT, (b) ALMA ACA, and (c) ALMA 12-m polarization data, with green crosses marking dense cores. To highlight regions with significant curvature, a 3$\sigma_{\rm curve}$ mask (0.0045, 0.024, and 0.165~arcsec$^{-1}$ for the JCMT, ACA, and 12-m data, respectively) is applied. The white curves mark the locations of the apparent U-shaped magnetic fields (as shown in \autoref{fig:pmap}) and coincide with regions of relatively high curvature. The green crosses label the dense cores identified in \citet{su21}.}
\label{fig:CurveMap}
\end{figure*}

The resulting curvature maps are shown in \autoref{fig:CurveMap}. A $3\sigma_{\rm curve}$ mask is applied to highlight only the statistically significant magnetic-field curvature structures. The U-shaped magnetic field morphologies identified in \autoref{fig:pmap}, marked with white curves, coincide with regions of relatively high curvature, particularly near the apex of each U-shaped feature. Additional regions of enhanced curvature are also visible in the 12-m curvature map surrounding the innermost protocluster region. These features are likely associated with the spiral-like structures converging toward the center rather than with U-shaped field morphologies.

\begin{figure}
\includegraphics[width=\columnwidth]{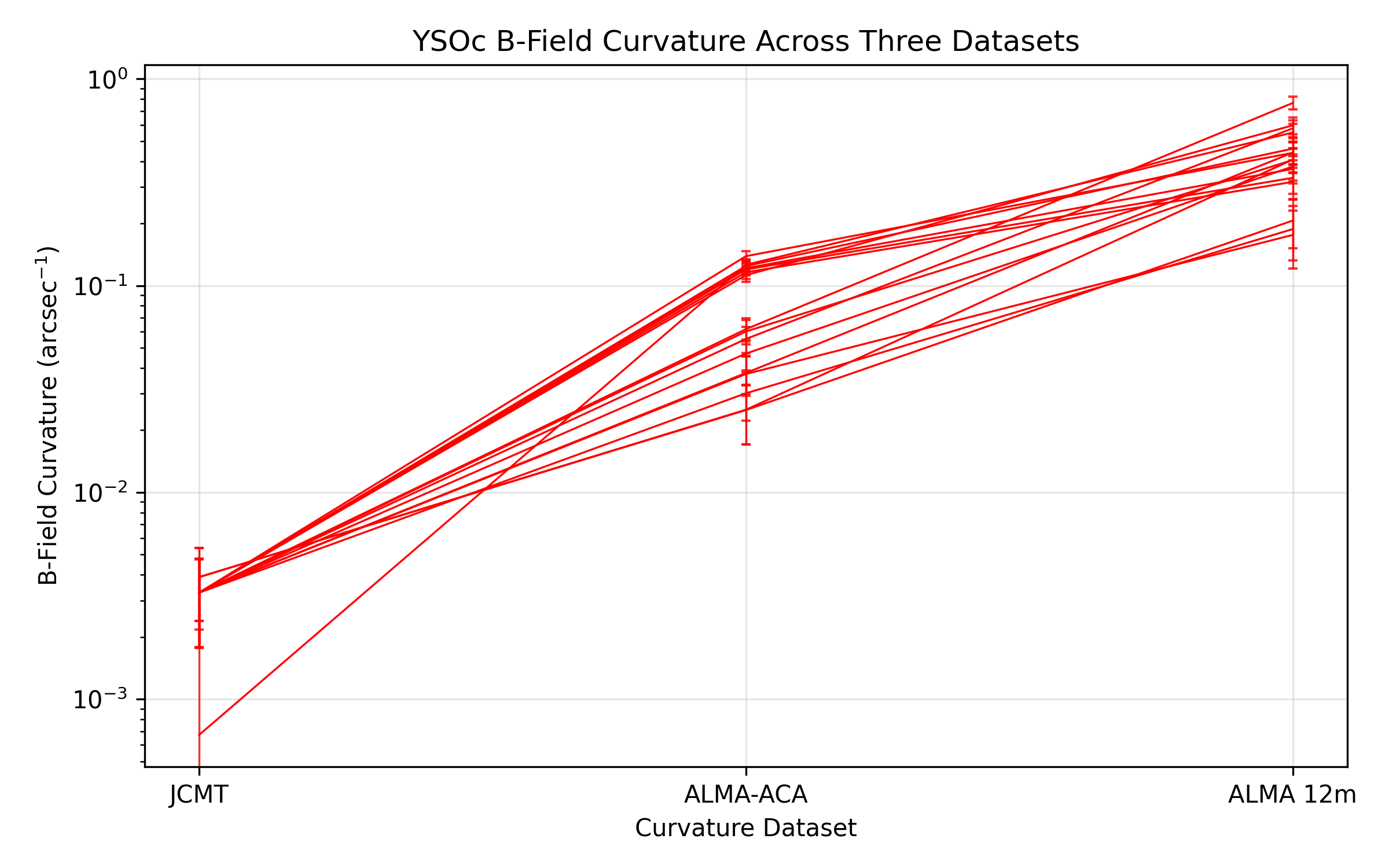}
\caption{Variation of magnetic field curvature ($\kappa$) over 
multiple scales towards individual cores: large scale (pc-scale converging filaments) -- intermediate scale (subparsec-scale colliding filaments) -- small scale (envelope-scale with flows and compressed layers).
$\kappa$ in most of the cores shows an order of magnitude increase from parsec- to subparsec-scale, and from subparsec- to envelope-scale.}\label{fig:multiCurve}
\end{figure}

\autoref{fig:multiCurve} presents the variation of magnetic-field curvature at the positions of dense cores identified by \citet{su21}. Cores without significant curvature on the ACA and 12-m scales are excluded, while the curvature estimates on the JCMT-scale are treated as upper limits. The results show an order-of-magnitude increase in magnetic-field curvature across these spatial scales, possibly driven by compression from accretion.

\subsection{Filament Identification}\label{sec:3.2}
To verify the filamentary morphology, we identified filaments with the Python package FilFinder \citep{filfinder} across multiple spatial scales. FilFinder first flattens the image using an arctangent transform and estimates the mean and standard deviation in the transformed space by fitting a log-normal brightness distribution. An adaptive threshold then creates a mask by requiring each central pixel in the smoothed image to exceed the median of its local neighborhood. This mask is combined with a global $3\sigma$ mask to exclude sub–noise features. Structures within the final mask are reduced to skeletons via a medial-axis transform.

\begin{figure*}
\includegraphics[width=\textwidth]{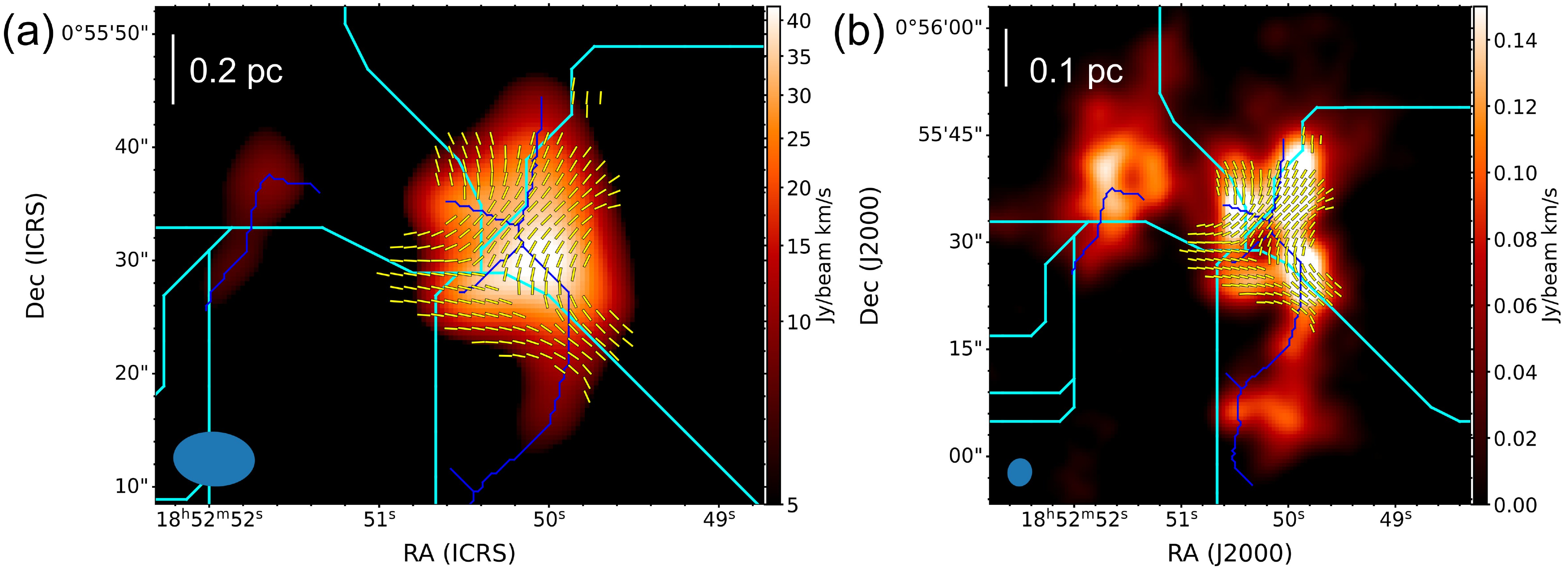}
\caption{Filaments identified at parsec (cyan) and subparsec (blue) scale, overlaid on the (a) ACA C$^{18}$O integrated intensity map and (b) the JVLA NH$_{3}$ integrated intensity map.}
\label{fig:ACA_Fi}
\end{figure*}

Since parsec-scale filaments have already been identified by \citet{wa20}, we focus here on the subparsec- and envelope-scales. For the subparsec-scale, we run FilFinder on the ACA C$^{18}$O integrated-intensity map rather than the continuum, as the continuum is dominated by the opaque hub, which obscures narrow filaments. A flatten percentage of $99.5\%$ is used to include pixels with intensity ranging within 95$\%$ of the peak value. The resulting skeletons (\autoref{fig:ACA_Fi}) overlaid on the C$^{18}$O and NH$_3$ moment-0 maps reveal two geometric families relative to the magnetic field: (i) filaments co-aligned with the U-shaped field, and (ii) filaments pointing toward the U-shape, likely linking to parsec-scale filaments. The NH$_3$ map traces filaments similar to those of C$^{18}$O, except for a central cavity likely caused by the locally reduced NH$_3$ abundance. The parsec-scale filaments identified by \citet{wa20}, with a threshold of 20~mJy/beam, from the JCMT 850~$\mu$m continuum are also shown in \autoref{fig:ACA_Fi}. A subset of these structures has orientations consistent with the ACA-scale filaments. A positional offset is visible, likely arising from the lower resolution and large-scale contamination in the JCMT data.

At the envelope scale, we apply FilFinder to the ALMA C-1 Band~6 continuum map. A flatten percentage of $95\%$ is used. The branches shorter than 2.5 arcsec are excluded. The identified skeletons (\autoref{fig:ACA_Fi}), shown with the Band~6 continuum and $^{13}$CS moment-0 maps, form spiral-like patterns that converge on the central massive protocluster. The $^{13}$CS emission broadly follows these filaments, though some branches are absent in $^{13}$CS, possibly due to spatially variable abundances (e.g., partial freeze-out onto grains or shock enhancements \citep{liu15}). For the subsequent kinematic analysis (Fig.~\autoref{fig:13CS}), we therefore select only filaments with clear $^{13}$CS detections. Comparing with the ACA-scale filaments, the envelope-scale filaments show consistent orientations toward the outer regions. Closer to the innermost region, the correspondence becomes less clear, likely reflecting the reduced molecule abundance. This trend is supported by the closer alignment between the ACA-scale filaments and the $^{13}$CS structures, compared to the continuum map (\autoref{fig:ACA_Fi}).  

\begin{figure*}
\includegraphics[width=\textwidth]{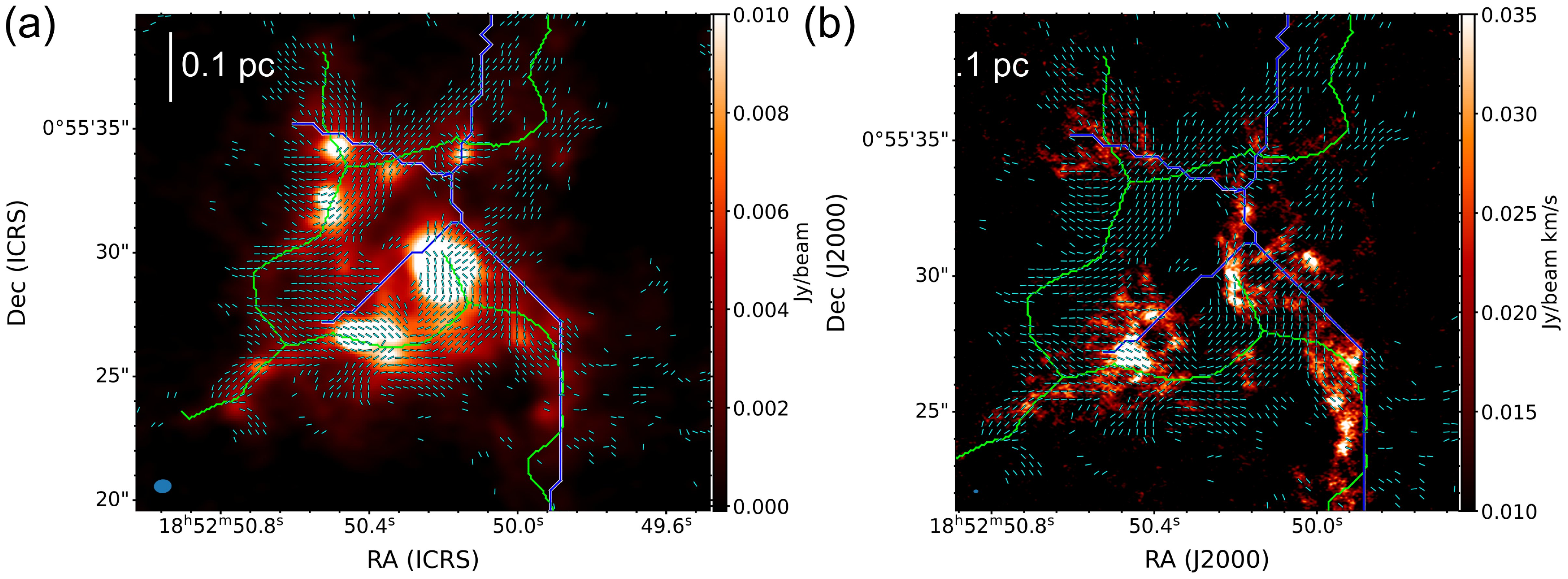}
\caption{(a) Filaments identified at subparsec (blue) and envelope-scale (green), overlaid on the (a) ALMA band 6 continuum and (b) the ALMA $^{13}CS$ integrated intensity map.}
\label{fig:12m_Fi}
\end{figure*}

\subsection{Relative Orientation Analysis}\label{sec:RO}
To quantify the alignment between the local magnetic field and filaments, we perform a local relative orientation analysis using the ACA and 12-m polarization data together with the identified filament structures. For each pixel along a filament, the local filament orientation ($\theta_{\rm fil}$) is estimated using the two adjacent pixels on either side, totaling five pixels, with a linear fit. The nearest polarization measurement within a distance of two beam sizes is then associated with each filament pixel. The two end pixels of each filament are excluded, as the local orientation is not well defined at the boundaries. 

\begin{figure*}
\includegraphics[width=\textwidth]{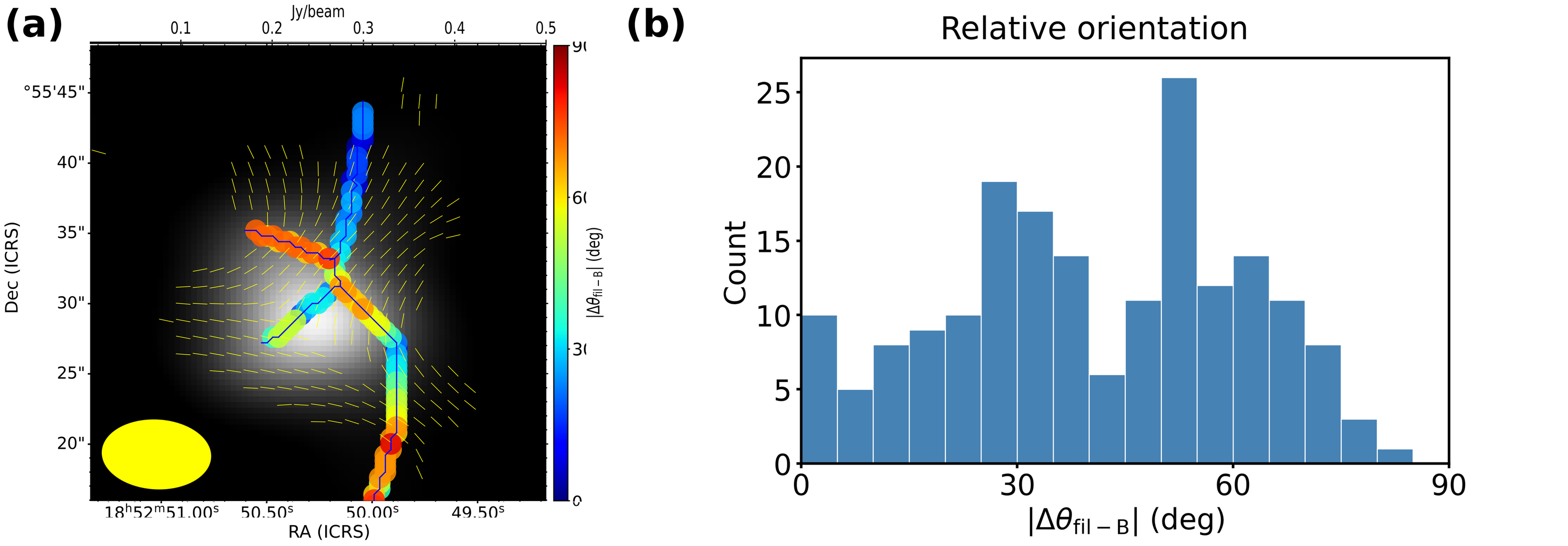}
\caption{Local relative orientation between the magnetic field and subparsec-scale filaments. (a) Relative orientations are indicated by the color-coded circles overlaid on the ACA continuum map. (b) Histogram of relative orientations across the map. Two distinct alignment modes (parallel-like and perpendicular-like) are identified in the NE–SW and NW–SE filaments.}
\label{fig:ACAro}
\end{figure*}

\begin{figure*}
\includegraphics[width=\textwidth]{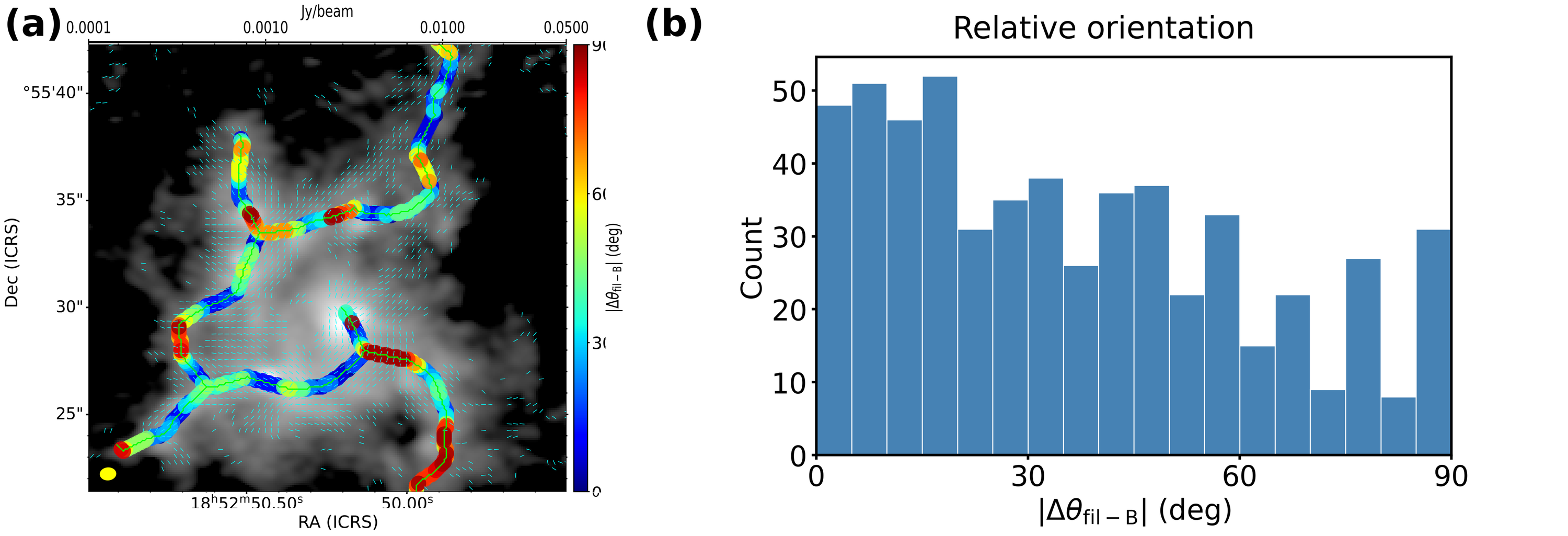}
\caption{Local relative orientation between the magnetic field and envelope-scale filaments. (a) Relative orientations are indicated by color-coded circles overlaid on the 12-m continuum map. (b) Histogram of relative orientations across the region. Most filaments exhibit a predominantly parallel-like alignment, although perpendicular-like orientations are still present in localized areas.}
\label{fig:ALMAro}
\end{figure*}

The relative orientation ($|\Delta\theta_{\rm fil-B}|$) between the local filament orientation and the magnetic field is computed and presented in \autoref{fig:ACAro} and \autoref{fig:ALMAro}, both as spatial maps and histograms to illustrate the distribution and overall trends.

On subparsec scales, the relative orientation confirms the presence of two alignment modes. The NE–SW filaments predominantly exhibit relative orientations greater than $60\degr$, whereas the NW–SE filaments tend to show relative orientations smaller than $45\degr$. This dichotomy is also reflected in the bimodal distribution seen in the histogram. We note that, due to projection effects and the properties of the tracers, deviations from perfectly parallel ($0\degr$) or perpendicular ($90\degr$) alignment are expected.

On envelope scales, the relative orientation shows a predominantly parallel-like alignment, as seen in both the spatial distribution and the histogram. This likely reflects a closer alignment of the magnetic field with the large-scale accretion flows. Nevertheless, regions with relative orientations exceeding $45\degr$ are still present, and these are spatially coincident with the U-shaped magnetic field structures shown in \autoref{fig:CurveMap}.

\subsection{Gas Kinematics}\label{sec:kinematics}
To investigate the interaction between gas kinematics and the observed U-shaped magnetic field structures, we make use of multi-scale molecular line observations. Given the complexity of the velocity structure, together with the limited resolution and sensitivity of the data, we focus on characterizing the overall velocity variation with multiple velocity ranges rather than attempting to resolve local velocity gradients. Complete channel maps of the line data are presented in \autoref{sec:appendix}.

\subsubsection{Parsec-Scale}
The IRAM 30-m C$^{18}$O (2–1) line data ($\theta_{beam}$= 0.4 pc) trace parsec-scale, moderate-density (n$_{H_2}\sim10^{-3}-10^{-4}$ cm$^{-3}$) gas, allowing comparison with the JCMT magnetic field map, shown in \autoref{fig:C18O}(a). To illustrate the overall velocity variation, we consider two velocity ranges associated with the two accreting filaments. Integrated intensity maps over 104–107 km~s$^{-1}$ (yellow) and 108–111 km~s$^{-1}$ (magenta) reveal velocity gradients along filaments (cyan lines, \citep{wa20}) converging toward the central hub where the massive protocluster locates (star symbol). The observed U-shaped magnetic field morphologies in the JCMT data are well-aligned with the accreting directions of these filaments. This suggests that the filaments are converging at the hub center, potentially triggering the formation of the massive protocluster, indicating gravitational inflows dragging the B-field directions.

\subsubsection{Subparsec-Scale}
\begin{figure*}
\includegraphics[width=\textwidth]{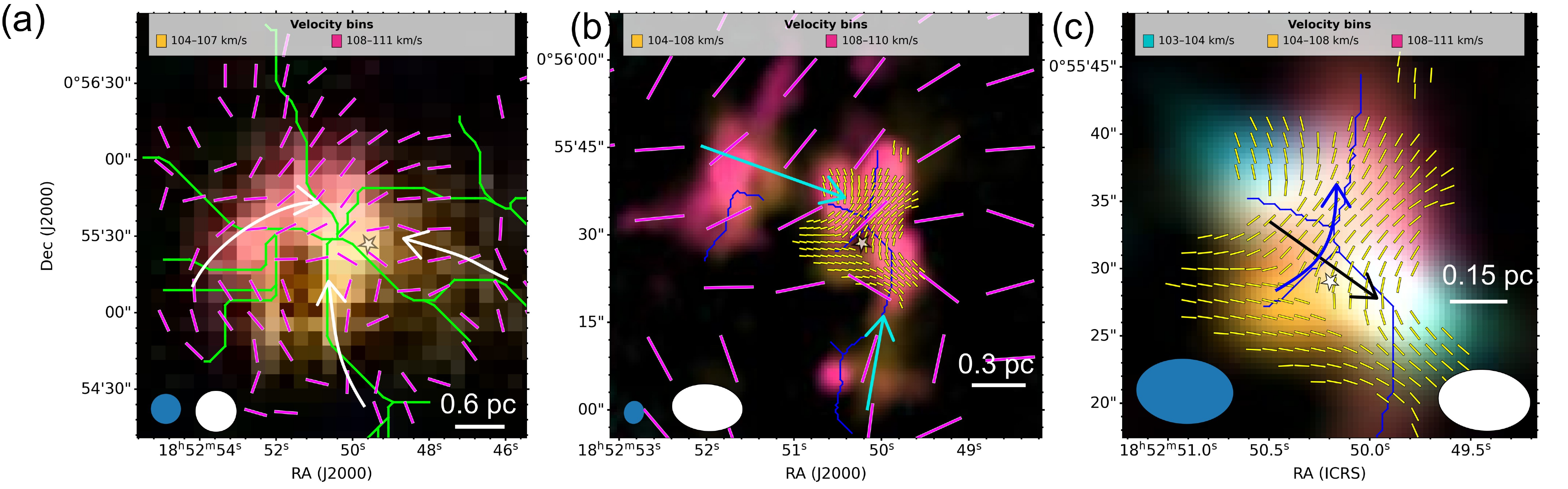}
\caption{Three color composite maps of C$^{18}$O (2-1) and NH$_{3}$ (1,1) integrated intensity across different velocity ranges. The blue and white beam are for velocity and polarization data, respectively. (a) JCMT polarization segments (magenta) overlaid on the IRAM 30-m C$^{18}$O (2–1) map, with identified filaments (green lines \citep{wa20}) converging toward the central hub (white star). (b) JCMT polarization segments (magenta) and ALMA ACA polarization segments (yellow) overlaid on the NH$_{3}$ (1,1) map. Several arc-like clumps and filaments converge from northeast and south toward the center (cyan arrows) following the similar direction as the parsec-scale filaments in (a), likely connecting to the subparsec-scale filaments (blue lines).
(c) 
ALMA ACA polarization segments (yellow) overlaid on the ACA C$^{18}$O (2–1) map. Clear velocity gradient are shown along the identified subparsec-scale filaments (blue lines): the cyan component traces inflows from the northeast and southeast (black arrow), as revealed in (a), that bend the U-shaped magnetic field, while the yellow and magenta components form an arc-like velocity gradient along the magnetic field (blue arrow). White star labels location of the protocluster. Animated channel maps with full velocity coverage are provided in \autoref{sec:appendix}.}
\label{fig:C18O}
\end{figure*}

\autoref{fig:C18O}(b) presents the JVLA NH$_3$ (1,1) data ($\theta_{beam}$= 0.14$\times$0.13 pc) with two velocity ranges highlighting the velocity variation along the filaments (108–111 km~s$^{-1}$, yellow) toward compact structures (108–110 km~s$^{-1}$). This map reveals filaments and arc-like clumps aligned with the local magnetic field and distributed along the large-scale filaments, indicating that these accreting filaments compress nearby fields and gas, ultimately channeling material into the central clump.

The ACA C$^{18}$O (2–1) line data ($\theta_{beam}$= 0.25$\times$0.15 pc)provide a detailed view of the gas kinematics at subparsec-scales and allow direct comparison with the ACA magnetic field morphology in \autoref{fig:C18O}(c). The integrated intensity maps reveal two distinct alignment modes between filaments (blue lines) and magnetic fields. Emission in the 103–104 km~s$^{-1}$ range (cyan) follows the direction of the parsec-scale filaments and aligns toward the peaks of the two U-shaped magnetic field structures on subparsec-scales, implying that this gas component may be inherited from the parsec-scale flow. In contrast, emission in the 104–108 km~s$^{-1}$ (yellow) and 108–111 km~s$^{-1}$ (magenta) ranges shows strong velocity gradients aligned with the curved U-shaped magnetic fields. These gradients are not evident on parsec-scale, suggesting that the structures likely formed as a result of filament–filament collisions.

\begin{figure*}
\includegraphics[width=\textwidth]{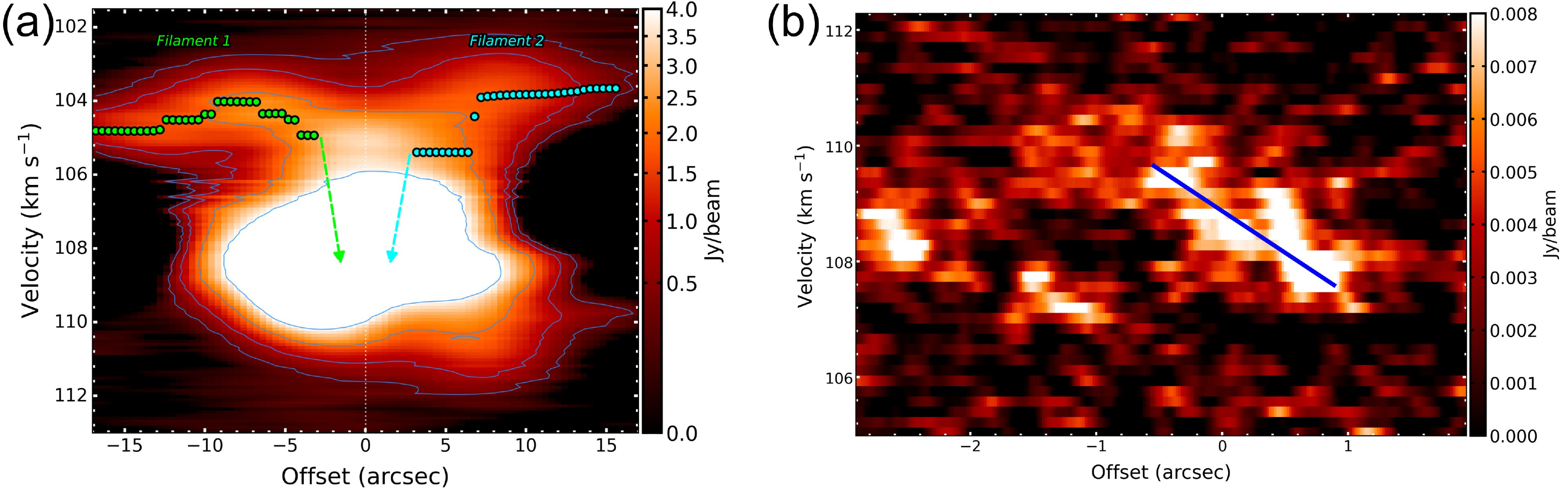}
\caption{(a) ACA C$^{18}$O position–velocity plot along the U-shaped feature (black arrow in \autoref{fig:C18O}c), showing filaments converging toward the brightest clump. The two filaments are represented by the intensity peaks (green and cyan points) within velocity between 101.5--105.5\kms at each offset until merged into the innermost $\pm$3\arcsec regions, forming a V-shaped feature. This V-shaped feature spans $\sim$2 \kms and corresponds to the cyan component in \autoref{fig:C18O}c, a signature often seen in cloud–cloud collisions. (b) 12-m array $^{13}$CS position–velocity plot along the U-shaped feature (green line in \autoref{fig:13CS}a), revealing a velocity gradient spanning $\sim$2 \kms along the U-shaped magnetic field.}
\label{fig:G33PV}
\end{figure*}

A position–velocity (PV) diagram along the direction of the blue component (\autoref{fig:G33PV}(a)) shows that multiple velocity components merge toward the central dense hub, forming a subparsec-scale V-shaped pattern associated with parsec-scale filaments, a signature commonly associated with cloud–cloud collisions \citep{ha15,ha15b}. To trace the kinematic structure of the two colliding filaments in the PV diagram, we identify intensity-peak positions as a function of spatial offset. For each spatial-offset column in the PV diagram, we extract the one-dimensional intensity profile and search for local maxima within the velocity range $101.5$--$105.5\ \mathrm{km\ s^{-1}}$ using the peak-finding algorithm of \texttt{scipy.signal.find\_peaks}. A candidate peak is accepted only if its intensity exceeds $0.3\ \mathrm{Jy\ beam^{-1}}$, its prominence above the local continuum is at least $0.4\ \mathrm{Jy\ beam^{-1}}$, and it is separated from any other accepted peak by a minimum of $3.5\ \mathrm{km\ s^{-1}}$. Peaks at negative offsets ($\leq 0\arcsec$) and positive offsets ($\geq 0\arcsec$) are assigned to Filament~1 and Filament~2, respectively; points within $\pm 3\arcsec$ of the phase centre are excluded to avoid contamination from the central clump. To illustrate the convergence of the two filaments toward the dense clump, dashed arrows are drawn from the innermost traced peak at $\pm 3\arcsec$ to the clump centroid.

\subsubsection{Envelope-Scale}
\begin{figure*}
\includegraphics[width=\textwidth]{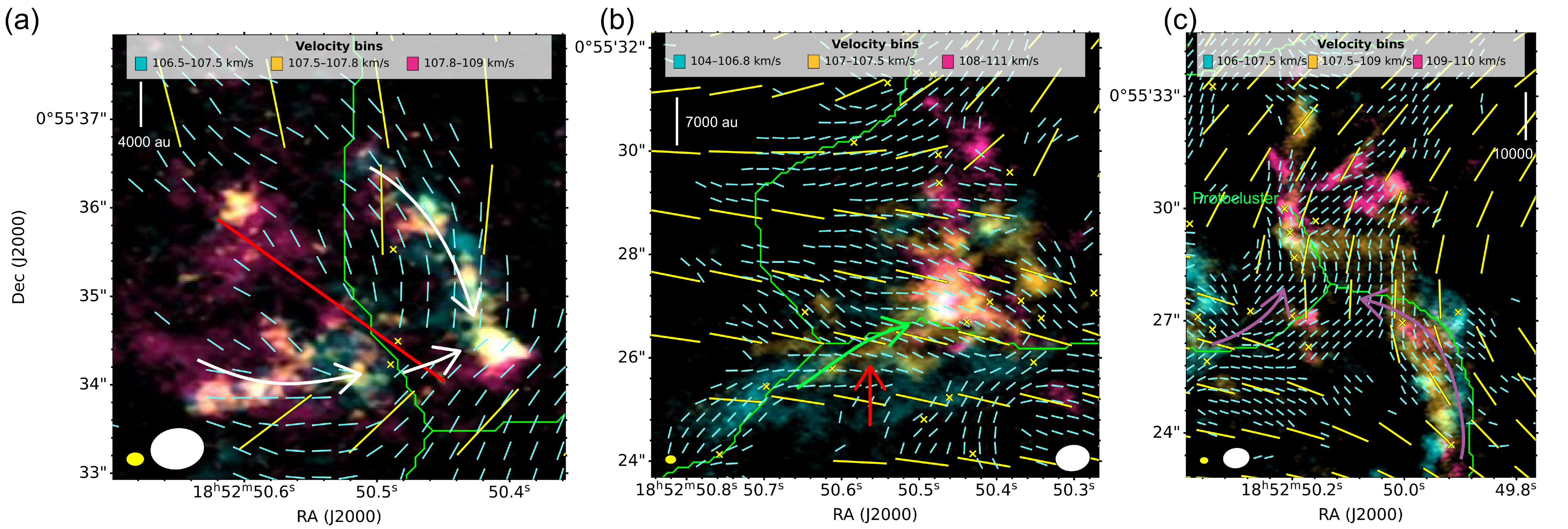}
\caption{ALMA 12-m (cyan) and ACA polarization (yellow) segments overlaid on three-color composite maps of $^{13}$CS (5--4) integrated intensity across selected velocity ranges. The yellow and white circles show the line and 12-m polarization data beam sizes. The green lines label the identified envelope-scale filaments.
(a) Several streamers from the north and northeast show a magenta--yellow--cyan velocity gradient following the U-shaped magnetic field (white arrow), fragmented toward multiple dense cores.
(b) The southeastern filament exhibits a velocity gradient across nearly parallel layers, from south (cyan) to north (yellow), following the direction of the curved field morphology (red arrow). This suggests that the field is shaped by large-scale layer compression. A cyan--yellow--magenta gradient is also seen from southeast to northwest along the field lines (green arrow), indicating that gas in the compressed layers is accreting toward the central clump along the magnetic field. 
(c) Filaments near the central protocluster display velocity gradients aligned with the magnetic field, becoming more spiral-like as infall and rotation dominate closer to the center. 
Yellow crosses mark the dense cores identified in \citet{su21}, which are distributed both within the filaments and along the compressed layers. An animated channel map with full velocity coverage is provided in \autoref{sec:appendix}.}
\label{fig:13CS}
\end{figure*}

To further probe the envelope-scale kinematics, we use high-resolution ($\theta_{beam}$=900$\times$1200 AU) ALMA 12-m array observations of the high-density gas tracer $^{13}$CS (5–4) \citep{liu15}, which are compared with the corresponding polarization data \autoref{fig:13CS}. Although large-scale flows are filtered out in the interferometric data, the $^{13}$CS line reveals numerous streamer-like structures on envelope scales, following the filaments identified in continuum (green lines). These streamers often exhibit velocity gradients aligned with local magnetic field orientations, converging toward the central region. Upon reaching a dense layer, the streamers, and the aligned magnetic fields, abruptly become perpendicular to the inflow direction. This transition suggests that the magnetic fields, strengthened and bent by large-scale accretion and collision, play a crucial role in guiding gas inflow toward the compact cores.

\subsection{Magnetic Strength Estimation}\label{sec:Bstr}
The Davis–Chandrasekhar–Fermi (DCF) method is commonly used to estimate magnetic field strengths in molecular clouds, under the assumption that the perturbation of the magnetic field morphology is primarily driven by turbulence. However, in environments where coherent motions dominate the field morphology, the turbulent perturbation is comparatively weak and difficult to disentangle from the coherent component. In such cases, curvature-based analyses of the magnetic field provide an alternative approach to estimate the field strength, particularly when the field morphology is shaped by coherent gas motions such as converging flows.

The observed alignment between the U-shaped magnetic field patterns and the velocity structures on (sub)parsec scales supports a scenario in which the magnetic fields are pinched by converging gas flows. Under this scenario, the observed magnetic field curvatures can be explained by the balance between ram pressure of the inflowing gas and magnetic tension, expressed as:

\begin{equation}\label{eq1}
\frac{\rho v_A^2}{R_c} = \frac{\rho v_l^2}{2L},
\end{equation}

or equivalently,

\begin{equation}\label{eq2}
\frac{R_c}{L} = \frac{2B^2}{\mu_0 \rho v_l^2},
\end{equation}

where $\rho$ is the gas density, $v_A$ is the Alfvén velocity, $R_c$ is the curvature radius of the magnetic field, $v_l$ is the inflow velocity, $L$ is the half-width of the flow, and $\mu_0$ is the magnetic permeability \citep{go18}. We note that, strictly speaking, the above equations define an upper limit for the magnetic field strength because the observed bent field morphologies point at a winning ram pressure.

As a representative example, we selected the northeastern U-shaped magnetic field structure shown in \autoref{fig:13CS}(a) to estimate the magnetic field strength at subparsec- and envelope-scale. This region shows a well-defined U-shape pattern in both the ACA polarization and the 12-m continuum maps, which is aligned with the direction of the local velocity gradient, indicating coherent gas inflow on both subparsec- and envelope-scales.

The average H$_2$ number density of the hub envelope, estimated from the ACA continuum data, is $7.4\times10^5$ cm$^{-3}$. In contrast, the average density of the dense cores identified in this region (ID:70) is $1.2 \times 10^8$ cm$^{-3}$ \citep{su21}, suggesting a two-order-of-magnitude increase from the subparsec-scale clump to the compact protostellar cores. The curvature $\kappa$, equivalent to $1/R_c$, of the U-shaped magnetic field structure increases from 0.12 to 0.60 arcsec$^{-1}$ from subparsec to envelope-scales, while the half-width of the structure decreases from 0.25 to 0.04 pc. The line-of-sight velocity difference across the U-shaped pattern remains consistent at 2.0 km~s$^{-1}$ on both scales. Given that this source is viewed nearly face-on \citep{liu12}, with flow motions and magnetic field morphology predominantly in the plane of the sky, we adopt an inclination angle of $10^\circ$ to deproject the velocity. We also assume that projection effects on magnetic field curvature are negligible.

The resulting estimates of the total magnetic field strength on the envelope and subparsec scales are 1.0\,mG and 14.1\,mG, respectively. This represents a substantial 
increase from the parsec-scale plane-of-sky magnetic field strength of $230 \pm 8~\mu\mathrm{G}$, derived from the JCMT data using the DCF method with an average density of $(1.0 \pm 0.1)\times10^{5}~\mathrm{cm^{-3}}$ \citep{wa21}. The corresponding Alfv\'en velocities at the envelope and subparsec scales are approximately 1.5 and 1.7~km\,s$^{-1}$, respectively. Given that the non-thermal linewidths in the ACA C$^{18}$O and ALMA $^{13}$CS data typically range from 1.5-2.5~km\,s$^{-1}$, the resulting Alfv\'enic Mach numbers lie in the trans- to super-Alfv\'enic regime. However, because the observed linewidths in such a dynamically active region may include contributions from collimated or systematic motions rather than purely turbulent broadening, the inferred Alfv\'enic Mach numbers should be interpreted as upper limits.

The primary source of uncertainty in our estimates stems from projection effects. For instance, varying the assumed inclination angle between $5^\circ$ and $10^\circ$ could alter the inferred velocity, and therefore the magnetic field strength, by a factor of two. Since a precise inclination angle is not available, we emphasize the relative change in magnetic field strength across spatial scales rather than the absolute values. If the subparsec-scale structures evolved directly from the larger-scale envelope, the projection effects at these two scales should remain similar and thus largely cancel out. 

\subsection{Magnetic Strength Scaling Factor}\label{sec3.4}
The magnetic field scaling relation ($B\propto n^{\alpha}$) provides insight into the role of magnetic fields in cloud evolution, as both field strength and density increase during gravitational collapse, and the degree of enhancement depends on whether the collapse proceeds against magnetic tension.. In weak-field scenarios, isotropic contraction leads to a scaling index $\alpha$ of $2/3$ \citep{me66}, as magnetic fields do not influence the collapse geometry. Conversely, a shallower index (e.g., $\sim$0.5 or lower) is expected when magnetic fields significantly regulate cloud collapse \citep{mo99}, since material preferentially contracts along field lines.

To statistically assess the role of magnetic fields across G33, we applied the same analysis to dense cores identified in this system \citep{su21} using high-resolution ALMA data with significant magnetic field curvature detected. At the locations of these dense cores, \autoref{fig:CurveMap} highlights variations in the local magnetic field curvature meeting a $3\sigma_{\kappa}$ detection threshold. Across the three datasets, the inferred curvature show order of magnitude increases toward smaller scales, from parsec to sub-parsec and, finally, to envelope scales, due to the evolving balance between magnetic tension and the ram pressure of the flow. Since many of these cores exhibit complex velocity structures in which the flow components cannot be cleanly disentangled, we adopt the same flow velocities shown in \autoref{fig:G33PV} for all sources. We note that the variation in velocity, typically less than a factor of two, is relatively minor compared to the order-of-magnitude differences in density and curvature.

We derived the gas densities at both the subparsec and envelope scales from the dust continuum data. At the subparsec scale, we used the ACA 1.3~mm continuum map (\autoref{fig:pmap}b) to calculate the total intensity and applied a Gaussian fit to the major clump to determine its radius. Since G33 hosts a massive and compact protocluster, we excluded the innermost compact region (radius $\sim$0.012~pc; see \autoref{fig:pmap}c) when computing the average density of the subparsec-scale clump. We adopted a dust temperature of 20~K and a dust opacity $\kappa = 0.005$~cm$^2$\,g$^{-1}$ at 1.3~mm \citep{hi83}. The resulting clump mass is 963~$M_\odot$ (excluding the protocluster), with a radius of $0.17\pm0.01$~pc, corresponding to an average H$_2$ density of $7.4\times10^5$~cm$^{-3}$, assuming a mean molecular weight of 2.8. At the envelope scale, we adopted the mass and radius from the catalog in \citet{su21}, based on ALMA Band~6 continuum data, and used these values to estimate the corresponding H$_2$ density, as listed in \autoref{tab2}. Note that the dominant source of uncertainty in the density estimates arises from the assumed dust opacity, which depends on the adopted dust model, rather than from observational errors. In this analysis, we focus only on the latter.

The projected widths of the U-shaped features were directly measured from the polarization maps, giving $\sim$0.5~pc at the envelope scale and $\sim$0.08~pc at the subparsec-scale. We note that these widths largely reflect the data resolution at the respective spatial scales, and thus appear relatively uniform across all U-shaped patterns in the map.

Using the above information, we estimated the magnetic-field strength scaling factor for all identified cores. We excluded four cores (IDs~43, 50, 53, and 56) located within the central ultracompact H\textsc{ii} region, where stellar feedback likely invalidates our assumptions, as well as two cores (IDs~0 and 44) lying outside the major structures with insufficient polarization detections to reliably determine the field curvature. We further exclude 15 samples without significant curvature ($<3\sigma_{\kappa}$).

\begin{figure}
\includegraphics[width=\columnwidth]{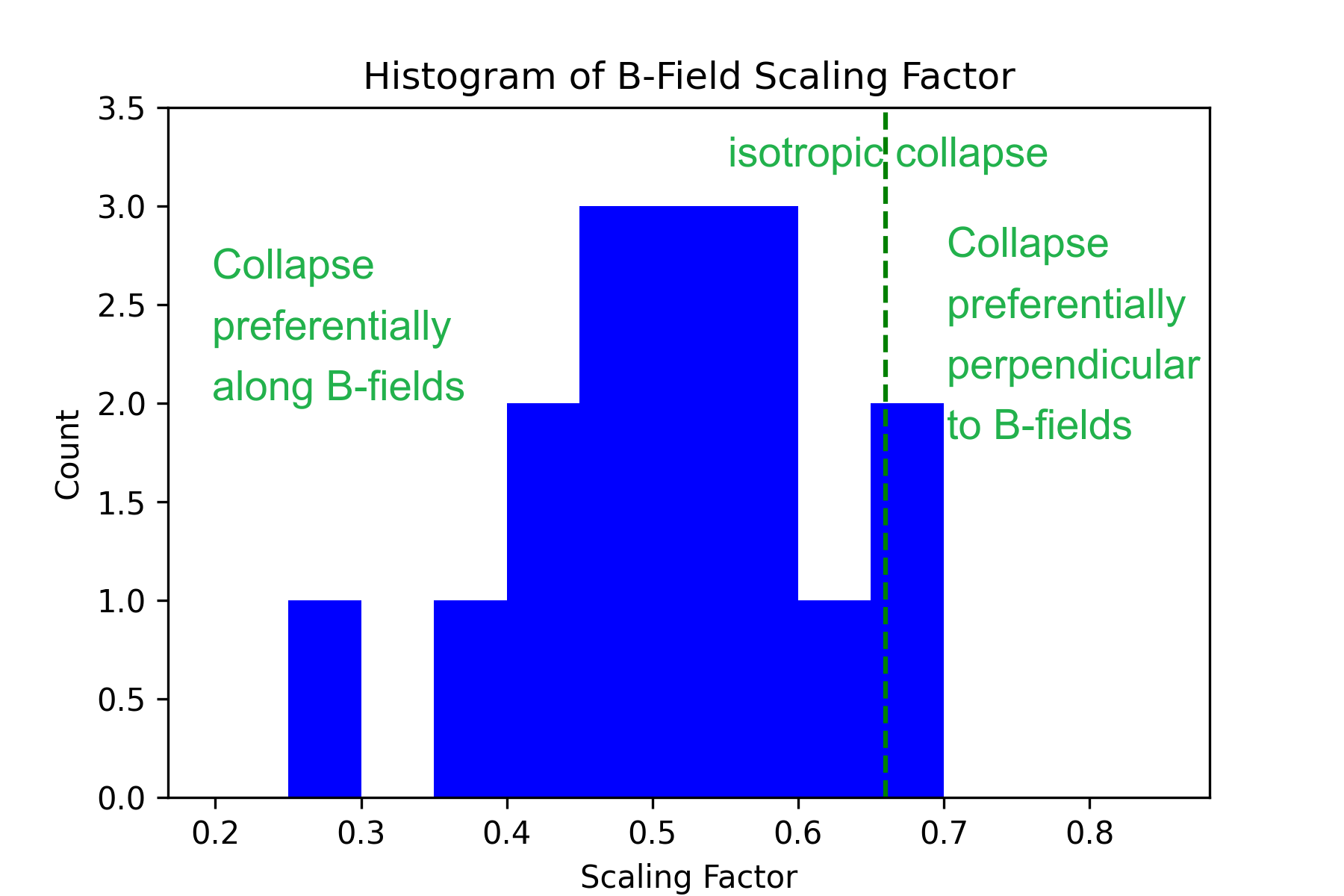}
\caption{Histogram of magnetic field strength scaling factors from dense cores. The peak is at 0.45-0.60 with a mean value of 0.50 and $\sigma=0.10$. The green dashed line indicates the factor of $2/3$ for an isotropic collapse.}\label{fig:hist_SF}
\end{figure}

The distribution of magnetic field scaling factors for the remaining 17 dense cores is shown in \autoref{fig:hist_SF}. We find a clear peak at scaling factors of 0.45--0.60 with a mean and standard deviation of 0.50$\pm$1.0. Our estimated index of 0.50, suggesting that magnetic fields are dynamically important in regulating gas kinematics from subparsec- to envelope-scales, consistent with the $^{13}$CS velocity gradients observed along filaments aligned with the magnetic field. We note that dynamically important magnetic fields are not inconsistent with our assumption of a local balance between ram pressure and magnetic tension. Such a balance can be established during the early stages when the U-shaped magnetic structures form, whereas the dynamical role of the magnetic field naturally evolves at later times as the system continues to evolve with time and varies spatially across the system.

\begin{deluxetable*}{ccccccccc}
\tablecaption{Magnetic Field Strength Scaling Factor\label{tab2}}
\renewcommand{\thetable}{\arabic{table}}
\tablehead{\colhead{ID} & \colhead{RA} & \colhead{Dec} & \colhead{Radius} & \colhead{Mass} & \colhead{$\rho$}  & \colhead{$\kappa_{\mathrm{ACA}}$} & \colhead{$\kappa_{\mathrm{12m}}$}& \colhead{Factor}\\
\colhead{} & \colhead{deg} & \colhead{deg} &\colhead{(pc)} &\colhead{($M_\odot$)} &\colhead{(cm$^{-3}$)} & \colhead{arcsec$^{-1}$} & \colhead{arcsec$^{-1}$} & \colhead{} }
\startdata
\hline
8 & 283.2101 & 0.92337 & 0.006 & $0.44 \pm 0.19$ & $(7.1 \pm 3.1)\times 10^{6}$ & 0.025 & 0.41 & $0.29 \pm 0.06$ \\
26 & 283.2098 & 0.92387 & 0.013 & $12.01 \pm 4.95$ & $(1.9 \pm 0.8)\times 10^{7}$ & 0.025 & 0.21 & $0.46 \pm 0.06$ \\
29 & 283.2083 & 0.92415 & 0.019 & $12.16 \pm 5.01$ & $(6.2 \pm 2.5)\times 10^{6}$ & 0.139 & 0.44 & $0.66 \pm 0.13$ \\
35 & 283.2098 & 0.92410 & 0.007 & $6.45 \pm 2.66$ & $(6.5 \pm 2.7)\times 10^{7}$ & 0.038 & 0.16 & $0.54 \pm 0.05$ \\
39 & 283.2099 & 0.92420 & 0.005 & $2.59 \pm 1.07$ & $(7.2 \pm 3.0)\times 10^{7}$ & 0.038 & 0.44 & $0.43 \pm 0.04$ \\
40 & 283.2095 & 0.92424 & 0.007 & $1.64 \pm 0.68$ & $(1.7 \pm 0.7)\times 10^{7}$ & 0.055 & 0.58 & $0.42 \pm 0.06$ \\
41 & 283.2100 & 0.92419 & 0.004 & $1.37 \pm 0.57$ & $(7.4 \pm 3.1)\times 10^{7}$ & 0.030 & 0.19 & $0.50 \pm 0.05$ \\
51 & 283.2103 & 0.92483 & 0.009 & $3.68 \pm 1.52$ & $(1.8 \pm 0.7)\times 10^{7}$ & 0.062 & 0.77 & $0.39 \pm 0.05$ \\
52 & 283.2099 & 0.92489 & 0.009 & $2.18 \pm 0.90$ & $(1.0 \pm 0.4)\times 10^{7}$ & 0.047 & 0.38 & $0.45 \pm 0.07$ \\
55 & 283.2103 & 0.92498 & 0.007 & $1.96 \pm 0.81$ & $(2.0 \pm 0.8)\times 10^{7}$ & 0.121 & 0.37 & $0.61 \pm 0.08$ \\
58 & 283.2106 & 0.92537 & 0.006 & $2.23 \pm 0.92$ & $(3.6 \pm 1.5)\times 10^{7}$ & 0.126 & 0.55 & $0.55 \pm 0.06$ \\
62 & 283.2103 & 0.92541 & 0.003 & $0.72 \pm 0.30$ & $(9.3 \pm 3.9)\times 10^{7}$ & 0.120 & 0.33 & $0.58 \pm 0.05$ \\
63 & 283.2105 & 0.92564 & 0.007 & $13.77 \pm 5.67$ & $(1.4 \pm 0.6)\times 10^{8}$ & 0.124 & 0.46 & $0.55 \pm 0.04$ \\
64 & 283.2098 & 0.92591 & 0.022 & $16.93 \pm 6.98$ & $(5.5 \pm 2.3)\times 10^{6}$ & 0.060 & 0.41 & $0.48 \pm 0.10$ \\
67 & 283.2079 & 0.92608 & 0.008 & $1.69 \pm 0.70$ & $(1.1 \pm 0.5)\times 10^{7}$ & 0.037 & 0.18 & $0.55 \pm 0.08$ \\
70 & 283.2104 & 0.92617 & 0.005 & $4.26 \pm 1.75$ & $(1.2 \pm 0.5)\times 10^{8}$ & 0.112 & 0.60 & $0.52 \pm 0.04$ \\
73 & 283.2104 & 0.92654 & 0.01 & $2.74 \pm 1.13$ & $(9.5 \pm 3.9)\times 10^{6}$ & 0.116 & 0.32 & $0.66 \pm 0.11$ \\
\enddata
\tablecomments{The source IDs, coordinates, radius, and mass are adopted from the catalog in \citet{su21}.}
\end{deluxetable*}

While our analysis offers a new approach
for estimating the magnetic field scaling factor, it is subject to several limitations due to the underlying simplified model. First, additional sources of kinematic energy, such as rotation, shocks from cloud–cloud collisions, and feedback from star formation, may be comparably important to accretion-driven motions in shaping the magnetic field morphology. 
Second, as density structures may keep evolving along magnetic field lines even after the field morphology is established, the present density structures may not always directly match the local magnetic field configuration. These factors can introduce systematic uncertainties and may explain the few outliers observed in our magnetic field scaling factor estimates (\autoref{fig:hist_SF}). Nevertheless, the majority of our sample shows consistent behavior, supporting the validity of our assumptions across most regions.

\subsection{Discussion}\label{sec:dis}
\begin{figure*}
\includegraphics[width=\textwidth]{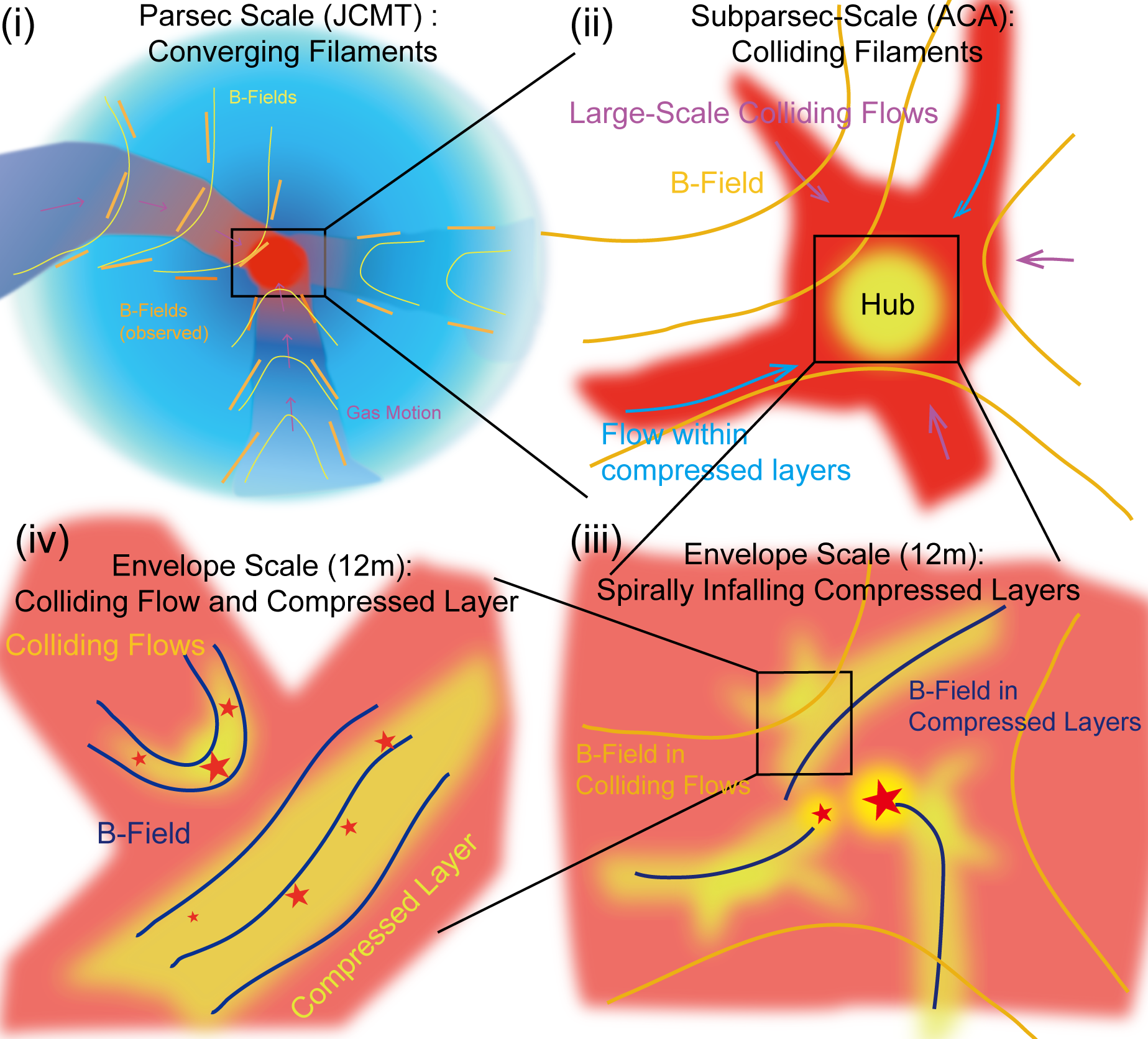}
\caption{Illustration of a possible evolutionary scenario in G33.92+0.11 from parsec to envelope scales via four scales.  
(i) On the parsec scale, three filaments converge toward the system, dragging the magnetic fields.  
(ii) On the sub-parsec scale, the three filaments collide, forming a massive hub and producing enhanced, and thus more curved U-shaped magnetic fields that point toward the hub.  
(iii) On the envelope scale, Three compressed layers originated from colliding flows, converge toward the system’s mass center along the magnetic fields, while spiraling inward due to the angular momentum inherited from the collision.
(iv) Zoom-in to these compressed layers, residual colliding flows with even furtther increased field curvature merge into the compressed layers. Dense cores may form either within the colliding flows with accretion guided by magnetic fields, or within the compressed layers, where fields are straightened and compressed.}
\label{fig:cartoon}
\end{figure*}

In order to explain the observed magnetic field morphology and scaling, along with the gas velocity structures, we propose a colliding flow scenario,
as illustrated in \autoref{fig:cartoon}.
On the parsec scales, large-scale filamentary flows appear to be converging toward the center of the system, as traced by the IRAM 30-m C$^{18}$O data. This convergence may be driven by global gravity \citep{wa22} or the large-scale expanding bubbles \citep{in15}. These filamentary flows also impact the parsec-scale magnetic field morphology, producing a converging pattern as revealed by the JCMT polarization observations.

On the subparsec-scales, these filamentary flows collide near the system's center, dragging magnetic fields along with them. This interaction gives rise to multiple U-shaped magnetic field structures, as seen in our ACA polarization data. Simultaneously, the collisions compress the gas, forming dense layers and enhancing the local magnetic field strength. These stronger magnetic fields likely play an important role in regulating local gas kinematics, which could explain the bright arc-like structures that align with the U-shaped magnetic fields in the ACA C$^{18}$O data (yellow and magenta components in \autoref{fig:C18O}c). In contrast, residuals of the parsec-scale filaments (cyan components in \autoref{fig:C18O}(c)) continue to flow inward, further pinching the magnetic field lines and increasing their curvature at smaller scales.

On the envelope scale, the U-shaped magnetic fields, strengthened by the colliding flows, are sufficiently strong to regulate the gas kinematics. Consequently, the $^{13}$CS data reveal streamers that are generally aligned with the local magnetic fields (\autoref{fig:13CS}a and c), consistent with a magnetic-field strength scaling factor of 0.44. The collision produces dense, compressed layers whose magnetic fields become more flattened as they are pressed from both sides. These layers then continue converging toward the massive center along the field lines.

Closer to the center, however, the angular momentum carried by the colliding flows, due to their non-zero impact parameters, twists the compressed layers and their associated magnetic fields, creating spiral-like structures aligned with the field lines (\autoref{fig:13CS}b). This spiraling pattern indicates that the magnetic-field morphology eventually becomes dominated by infall and rotational motions. The spirals converge toward the system’s mass center, forming a hub–filament structure that hosts a massive protocluster (\autoref{fig:13CS}b).

Dense cores are found in both the streamer regions, where the U-shaped magnetic-field patterns remain evident, and the central hub, where the streamers and magnetic fields gradually transition into spiral structures. The coexistence of these two distinct environments may lead to different evolutionary pathways, as the balance between magnetic and kinematic energy appears to reverse between the two environments.

Converging flows are a common phenomenon in theoretical studies of molecular cloud and star formation \citep[e.g.,][]{ha01,he08,va11}, and can be driven by a variety of mechanisms, including turbulence \citep[e.g.,][]{pa20}, expanding feedback-driven bubbles \citep[e.g.,][]{in12}, and global gravitational collapse \citep{va19}. These processes can naturally lead to the formation of filamentary structures and hub–filament systems \citep[e.g.,][]{ch14,go14,va19}. Recent observational studies further suggest that filament–filament merging or collisions may play an important role in triggering and regulating the evolution of hub–filament systems \citep[e.g.,][]{pi20, ku20,de24,ma25}. However, dedicated theoretical and numerical studies of filament–filament collisions remain relatively limited \citep[e.g.,][]{ho21,ka23,ka24}. In this work, our observationally motivated scenario provides a detailed view of how large-scale converging filaments influence the formation and evolution of hub–filament systems, and drive star formation across a range of environments, offering a strong empirical foundation for future theoretical investigations.

\section{Conclusions}\label{sec:con}
This study utilizes the multiscale JCMT POL-2, ALMA, and ACA continuum polarization observations to investigate the hub-filament system G33. An organized but complex magnetic field morphology is uncovered over parsec to 4000-au scales with the following main outcome.

\begin{itemize}
    \item {U-shaped magnetic-field morphologies are observed consistently across all spatial scales probed in this study. On parsec to subparsec scales, the apex of the U-shaped field is preferentially aligned with the large-scale gas-flow direction, while on smaller scales (subparsec to $\sim$4000 au) the gas velocity gradients become aligned with the local magnetic-field orientation. This transition suggests a scale-dependent coupling in which large-scale flows shape the magnetic-field geometry, the magnetic field regulates gas motions on smaller scales, and gravity ultimately dominates in the innermost region.
    
    \item ACA polarization observations reveal three U-shaped magnetic-field components converging toward the central hub of G33, each associated with a parsec-scale filament traced by molecular-line data. Position–velocity diagrams across these structures show characteristic V-shaped features over subparsec scales commonly seen in cloud–cloud collision systems, supporting a scenario in which massive star formation in G33 is triggered by multiple converging and colliding gas flows.

    \item Envelope-scale magnetic fields traced by high-resolution ALMA polarization data appear to be partially inherited from the subparsec-scale U-shaped morphology. On these scales, the magnetic field shows increased curvature within converging filaments, a more uniform configuration at the junctions of U-shaped components, and spiral patterns toward the massive central region, consistent with the influence of local condensation, compression, and rotationally infalling motions. Dense cores are found in all three regimes, indicating diverse initial conditions for star formation.

    \item We introduce a novel method to estimate magnetic-field strength and its scaling across multiple spatial scales using magnetic-field curvature, assuming a balance between magnetic tension and gas ram pressure during the formation of the U-shaped morphology. This approach yields a magnetic-field strength scaling of $0.50 \pm 0.10$ from sub-parsec to $\sim$4000 au scales, indicating that magnetic fields remain dynamically important and consistent with the observed alignment between velocity gradients and magnetic fields.
    
    \item We propose a unified scenario in which the collision of converging parsec-scale filaments establishes the global magnetic-field morphology, while compression-enhanced magnetic fields on smaller scales become sufficiently strong to influence core formation. In this picture, young stars may form within converging flows or in compressed layers, potentially with distinct initial conditions.}
\end{itemize}

\appendix
\section{Molecular Line Channel Maps}\label{sec:appendix}
\begin{figure}
\plotone{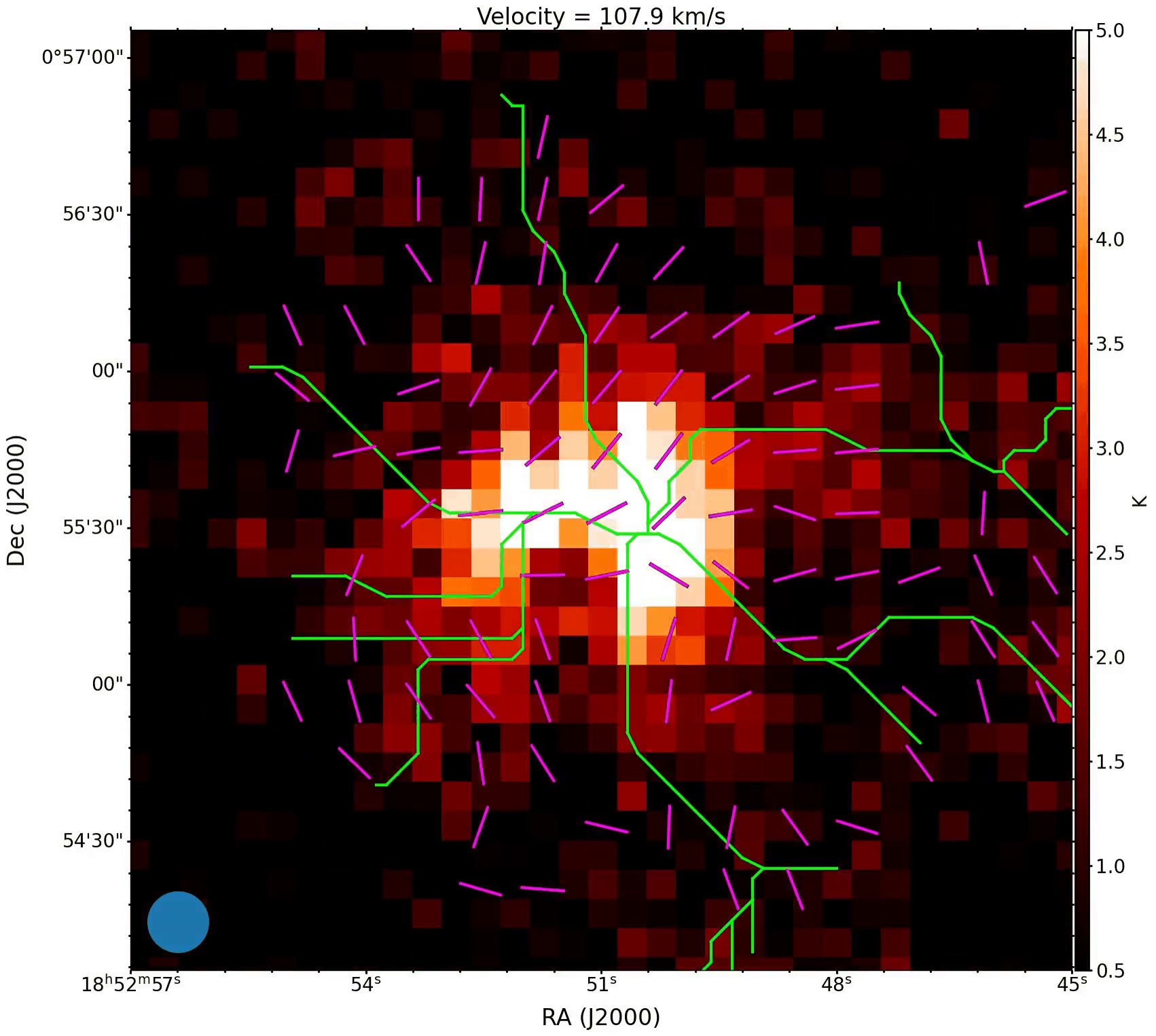}
\caption{Complete channel map for IRAM C$^{18}$O (2--1) data with POL-2 polarization segments (magenta segments) overlaid on the identified filaments (green). The ancillary movie steps through channel maps between 104.6 and 111.1 \kms\ and has a real-time duration of 9 sec. The static figure corresponds to one representative channel from the full animation, and a 2-color integrated intensity map highlighting the major velocity variation is shown in \autoref{fig:C18O}(a).  \label{fig:IRAMmovie}}
\end{figure}

\begin{figure}
\plotone{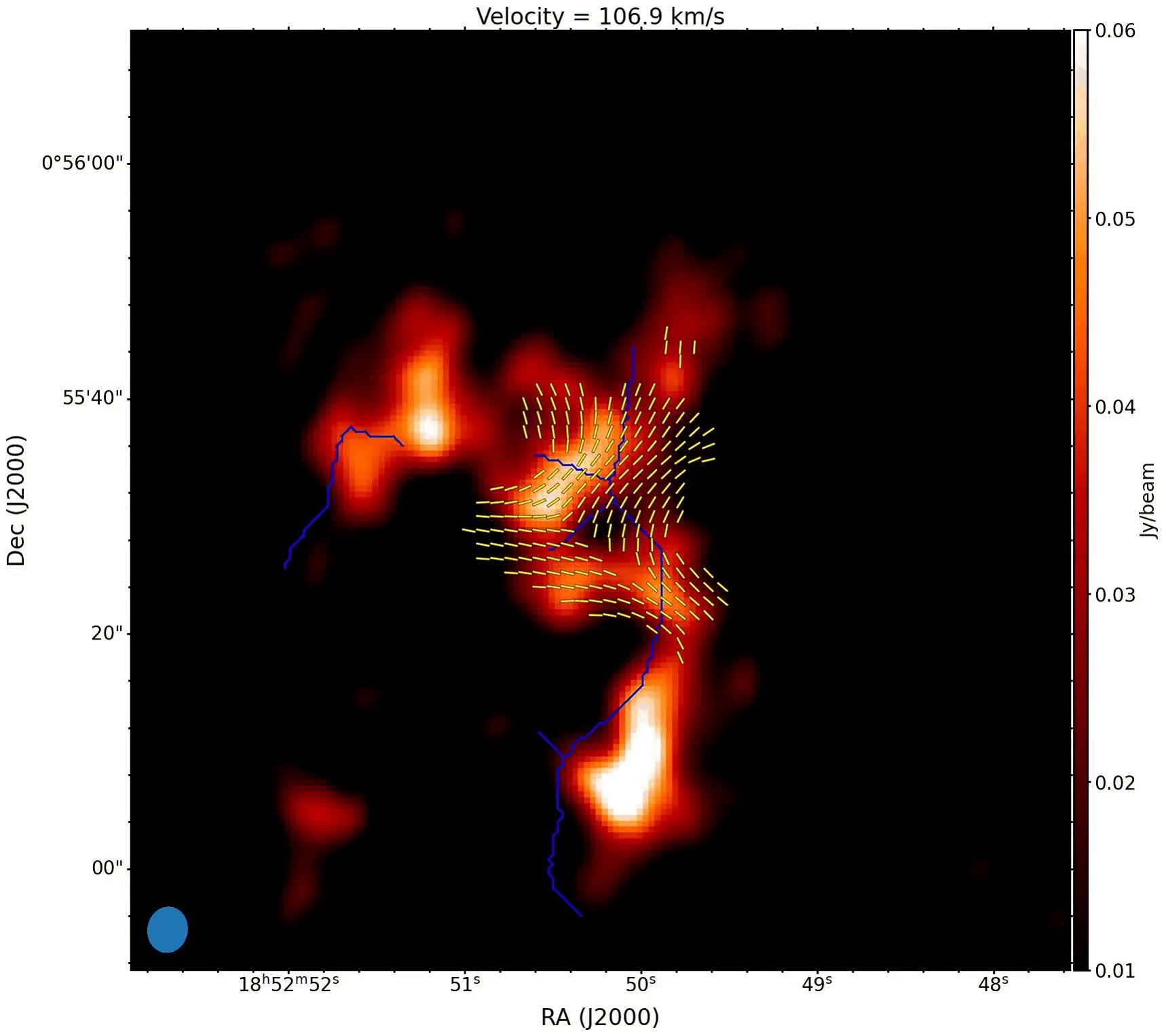}
\caption{Complete channel map for the JVLA NH$_{3}$ (1,1) line data (main hyperfine component only) with ACA polarization segments (yellow) and the identified filaments (blue) overlaid. The ancillary movie steps through channel maps between 104.4 and 110.3 \kms\ and has a real-time duration of 11 sec. The static figure corresponds to one representative channel from the full animation, and a 2-color integrated intensity map highlighting the major velocity variation is shown in \autoref{fig:C18O}(b). \label{fig:VLAmovie}}
\end{figure}

\begin{figure}
\plotone{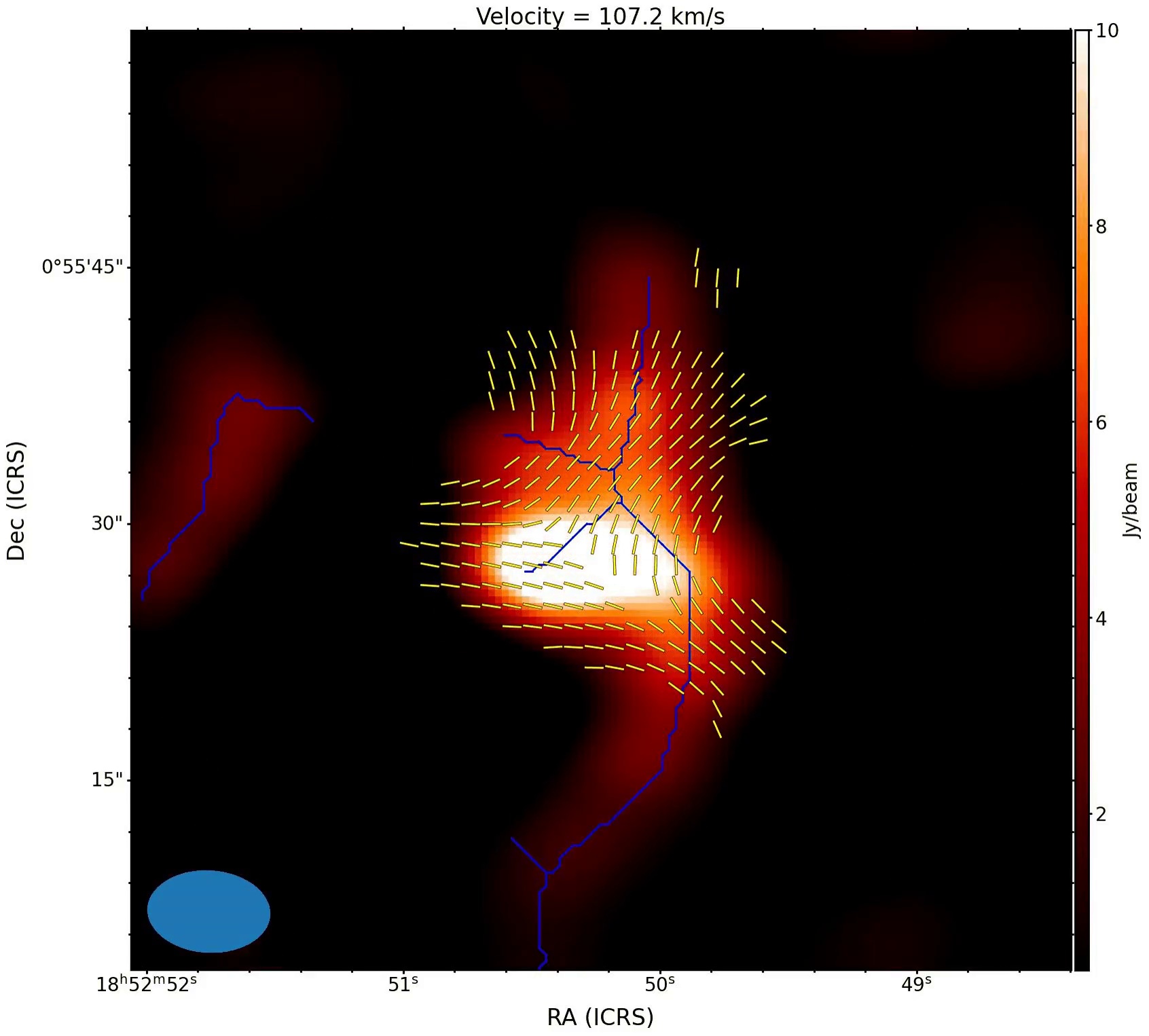}
\caption{Complete channel map for the ACA C$^{18}$O (2--1) data with ACA polarization segments (yellow) and identified filaments (blue) overlaid. The ancillary movie steps through channel maps between 100.9 and 112.7 \kms\ and has a real-time duration of 14 sec. The static figure corresponds to one representative channel from the full animation, and a 3-color integrated intensity map highlighting the major velocity variation is shown in \autoref{fig:C18O}(c). \label{fig:ACAmovie}}
\end{figure}

\begin{figure}
\plotone{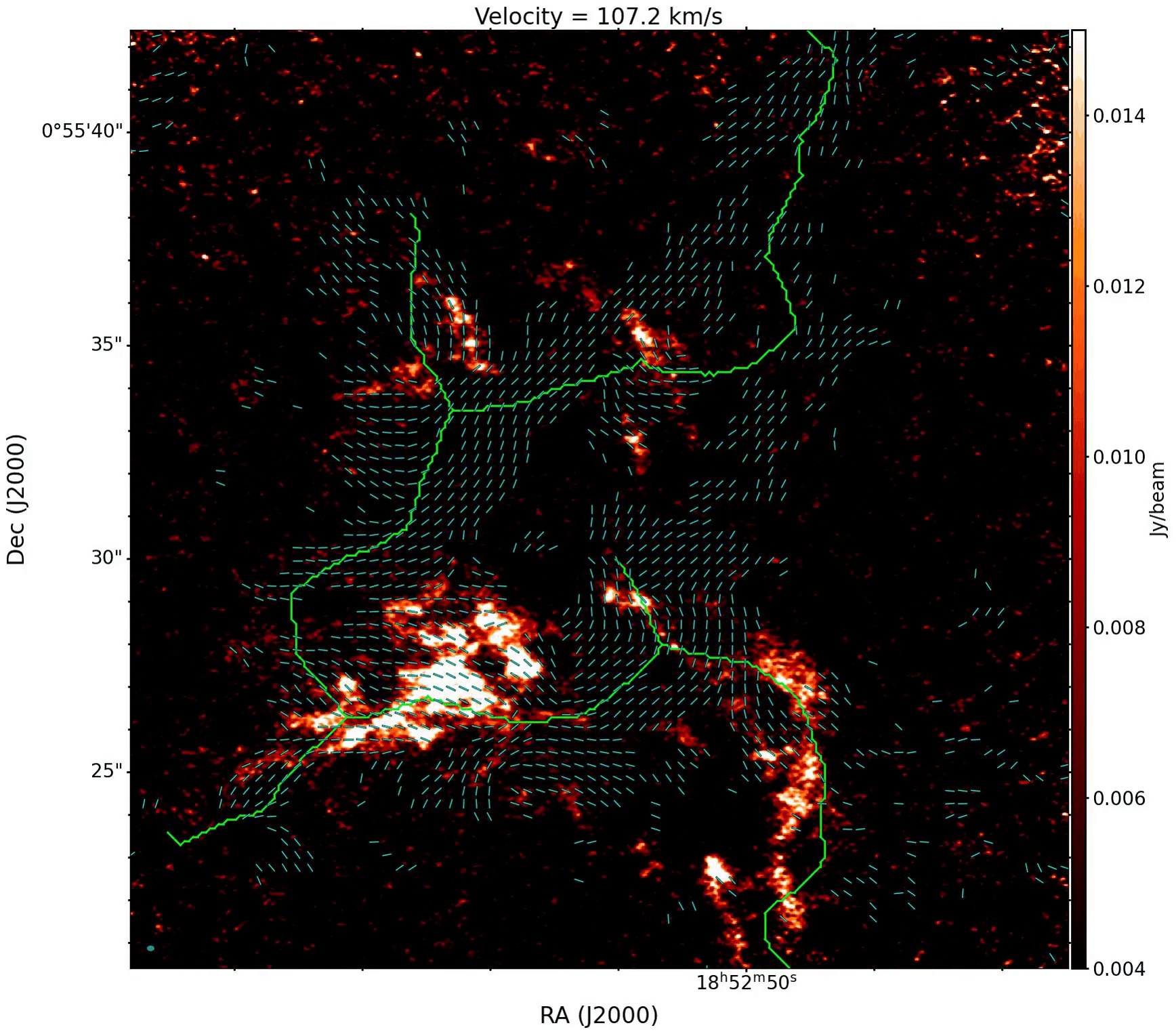}
\caption{Complete channel map for ALMA $^{13}$CS (5--4) data with ALMA 12M polarization segments (cyan segments) overlaid on the identified filaments (green). The ancillary movie steps through channel maps between 103.8 and 110.6 \kms\ and has a real-time duration of 12 sec. The static figure corresponds to one representative channel from the full animation, and a 3-color integrated intensity map highlighting the major velocity variation is shown in \autoref{fig:13CS}. \label{fig:13CSmovie}}
\end{figure}

In this section, we present representative channel maps for the IRAM C$^{18}$O, NH$_{3}$, ACA C$^{18}$O, and $^{13}$CS (5--4) data in \autoref{fig:IRAMmovie}, \autoref{fig:VLAmovie}, \autoref{fig:ACAmovie}, and \autoref{fig:13CSmovie}, respectively. The corresponding full channel-map movies are provided as ancillary files.

\acknowledgments
This paper makes use of the following ALMA data: ADS/JAO.ALMA\#2022.1.01482.S and 2023.1.01004.S. ALMA is a partnership of ESO (representing its member states), NSF (USA) and NINS (Japan), together with NRC (Canada), NSTC and ASIAA (Taiwan), and KASI (Republic of Korea), in cooperation with the Republic of Chile. The Joint ALMA Observatory is operated by ESO, AUI/NRAO and NAOJ. The National Radio Astronomy Observatory is a facility of the National Science Foundation operated under cooperative agreement by Associated Universities, Inc. We thank the NA and EA ALMA regional center to process the raw ALMA data. These observations were obtained by the James Clerk Maxwell Telescope, operated by the East Asian Observatory on behalf of Academia Sinica Institute of Astronomy and Astrophysics and the National Astronomical Research Institute of Thailand. Additional funding support is provided by the Science and Technology Facilities Council of the United Kingdom and participating universities and organizations in the United Kingdom and Canada. This work makes use of the imager and gildas software to reduce and analyze the data (See https://imager.oasu.u-bordeaux.fr and http://www.iram.fr/IRAMFR/GILDAS). 
PMK acknowledges support from the National Science and Technology Council (NSTC) in Taiwan 
through grants NSTC 112-2112-M-001-049-, NSTC 113-2112-M-001-016-, and NSTC 114-2112-M-001-007-.
H.B.L. is supported by NSTC with Grant No. 113-2112-M-110-022-MY3. H.B.L. wrote the IRAM proposal and calibrated the IRAM data, and the observations are performed by Roberto Galv\'{a}n-Madrid. H.B.L. and Y.L. conducted and calibrated the JVLA NH$_{3}$ observations. 
V.J.M.L.G. acknowledges support by the Spanish program Unidad de Excelencia María de Maeztu CEX2020-001058-M, financed by MCIN/AEI/10.13039/501100011033, and by the MaX-CSIC Excellence Award MaX4-SOMMA-ICE.
V.J.M.L.G. acknowledges support by the European Research Council (ERC) under the European Union’s Horizon 2020 research and innovation program (grant agreement No. 101098309 - PEBBLES).

\facilities{JCMT, ALMA, JVLA}
\software{Aplpy \citep{aplpy2012,aplpy2019}, Astropy \citep{astropy2013,astropy2018,astropy2022}, NumPy \citep{numpy}, SciPy \citep{scipy}, CASA \citep{casa}, GILDAS-IMAGER \citep{pe05}}

\bibliography{main}{}
\bibliographystyle{aasjournal}
\end{document}